\documentclass{aa}  
\usepackage{txfonts}
\usepackage{hyperref} 

\hypersetup{bookmarks=true,colorlinks=true,linkcolor=blue,citecolor=blue,urlcolor=blue}

\usepackage{supertabular}

\usepackage{array} 
\usepackage{tabularx}

\def\mps{\hbox{ms$^{-1}$}}
\def\kmps{\hbox{kms$^{-1}$}}

\def\mystar{GJ\,676A}
\def\mystarB{GJ\,676B}
\newcommand{\teta}{\boldsymbol{\theta}}

\defcitealias{Anglada-Escude:2012aa}{AT12}

\begin{document} 
   \title{The mass of planet \mystar \,b from ground-based astrometry\thanks{\textit{Based on observations collected at the European Organisation for Astronomical Research in the Southern Hemisphere under ESO programmes 385.C-0416 (A,B), 086.C-0515(A), 089.C-0115(D,E), 072.C-0488(E), 180.C-0886(A), 183.C-0437(A), 085.C-0019(A), 091.C-0034(A), 095.C-0551(A), 096.C-0460(A).}}}
   \subtitle{A planetary system with two mature gas giants suitable for direct imaging}

\author{J. Sahlmann\inst{1}\fnmsep\thanks{ESA Research Fellow}		
\and P. F. Lazorenko\inst{2}
\and D. S\'egransan\inst{3}
\and N. Astudillo-Defru\inst{3}
\and X. Bonfils\inst{4}
\and X. Delfosse\inst{4}
\and T. Forveille\inst{4} 
\and J.~Hagelberg\inst{5}
\and G. Lo Curto\inst{6}
\and F. Pepe\inst{3}
\and D. Queloz\inst{3,7} 
\and S. Udry\inst{3}
\and N.\,T. Zimmerman\inst{8}
}	

\institute{
European Space Agency, Space Telescope Science Institute, 3700 San Martin Drive, Baltimore, MD 21218, USA\\
\email{johannes.sahlmann@esa.int}
\and Main Astronomical Observatory, National Academy of Sciences of the Ukraine, Zabolotnogo 27, 03680 Kyiv, Ukraine
\and Observatoire de Gen\`eve, Universit\'e de Gen\`eve, 51 Chemin Des Maillettes, 1290 Versoix, Switzerland
\and Universit\'e Grenoble Alpes, CNRS, IPAG, 38000 Grenoble, France 
\and Institute for Astronomy, University of Hawai'i, 2680 Woodlawn Drive, Honolulu, HI 96822, USA
\and European Southern Observatory, Casilla 19001, Santiago, Chile
\and University of Cambridge, Cavendish Laboratory, J J Thomson Avenue, Cambridge, CB3 0HE, UK	
\and Space Telescope Science Institute, 3700 San Martin Drive, Baltimore, MD 21218, USA}
			
\date{Received 4 May 2016 / Accepted 29 July 2016} 

\abstract
{\mystar\ is an M0 dwarf hosting both gas-giant and super-Earth-type planets discovered with radial-velocity measurements. Using FORS2/VLT, we obtained position measurements of the star in the plane of the sky that tightly constrain its astrometric reflex motion caused by the super-Jupiter planet `b` in a 1052-day orbit. This allows us to determine the mass of this planet to $M_\mathrm{b} = 6.7^{+1.8}_{-1.5}\,M_\mathrm{J}$, which is $\sim$40 \% higher than the minimum mass inferred from the radial-velocity orbit. Using new HARPS radial-velocity measurements, we {improve upon the orbital parameters of the inner low-mass planets `d` and `e` and we} determine the orbital period of the outer giant planet `c` to $P_\mathrm{c}=7340$ days under the assumption of a circular orbit. {The preliminary minimum mass of  planet `c` is} $M_\mathrm{c} \sin i = 6.8\,M_\mathrm{J}$ {with an upper limit of $\sim$$39\,M_\mathrm{J}$ that we set using NACO/VLT high-contrast imaging}. We also determine precise parallaxes and relative proper motions for both \mystar\ and its wide M3 companion \mystarB. Despite the probably mature age of the system, the masses and projected separations ($\sim$0\farcs1 -- 0\farcs4) of planets `b` and `c` make them promising targets for direct imaging with future instruments in space and on extremely large telescopes. In particular, we estimate that \mystar \,b and \mystar \,c are promising targets for directly detecting their reflected light with the WFIRST space mission. Our study demonstrates the synergy of radial-velocity and astrometric surveys that is necessary to identify the best targets for such a mission.} 
\maketitle

\section{Introduction}

The discovery and characterisation of extrasolar planets is progressing at a staggering pace, fueled by new instrumentation and data analysis methods. Main sequence low-mass stars, the M dwarfs, represent an important target sample because they are the most numerous stars in the Galaxy and host a large number of small planets \citep{Bonfils:2013aa, Dressing:2013aa}. Giant planets around M dwarfs, however, are found to have a low occurrence compared to their counterparts around Sun-like stars {\citep[e.g.][]{Endl:2006uq, Cumming:2008lr, Bonfils:2013aa}}, which is an expected outcome of the core accretion scenario for planet formation \citep{Laughlin:2004uq}.

One example for a low-mass star harbouring giant planets is the M0 dwarf \href{http://simbad.u-strasbg.fr/simbad/sim-basic?Ident=GJ676A&submit=SIMBAD+search}{\mystar}, which is part of a wide ($\sim$800 AU) binary system of M-dwarfs located at a distance of $\sim$17 pc from the Sun. Using radial-velocity monitoring, \cite{Forveille:2011lr} discovered a giant planet (planet `b`) around \mystar\ with minimum mass $M \sin i=4.9\,M_\mathrm{J}$ and an orbital period of 2.9 years. \cite{Forveille:2011lr} also discovered an additional radial-velocity drift that could not be explained by the wide companion \mystarB, but required the presence of a second outer companion to \mystar. Then \citet{Anglada-Escude:2012aa} {(hereafter \citetalias{Anglada-Escude:2012aa})} reported the detection of two additional super-Earth planets in short-period orbits (planets `d` and `e`) and confirmed the presence of the outer companion, probably a second gas giant (planet `c`).

\mystar\ thus represents a rare case of a planetary system with inner super-Earths and outer gas-giant planets around an M dwarf, a configuration reminiscent of our Solar System. It appears that such systems typically are difficult to form \citep[e.g.][]{Raymond:2008aa}, yet other more compact examples have been found, e.g.\ around the M dwarf \object{GJ\, 876} \citep{Rivera:2010aa} and the two \emph{Kepler} transiting systems \object{KIC\,11442793} \citep{Cabrera:2014aa} and \object{KOI\,435} \citep{Ofir:2013aa}. 

Here, we present new astrometric, radial velocity, and high-contrast imaging observations that allow us to better characterise the planetary system around \mystar.

\section{Observations and data reduction}
Upon the discovery of planet \mystar\,b, we initiated an astrometric program to measure the star's orbital reflex motion caused by the planet, which yields an accurate planet mass measurement by determining the orbital inclination. The minimum semimajor axis of \mystar's barycentric orbit caused by planet `b` is $\sim$0.7 milli-arcsecond (mas), thus detectable with high-precision ground-based astrometry that reaches a per-epoch precision of order 0.1 mas \citep{Lazorenko:2011lr, Sahlmann:2014aa}.

\subsection{FORS2/VLT astrometry}\label{sec:fors2ax}
We obtained optical images of \mystar\ with the FORS2 camera \citep{Appenzeller:1998lr} installed at the Very Large Telescope (VLT) of the European Southern Observatory (ESO) between April 2010 and August 2012. The instrument setup and observation strategy, e.g.\ obtaining several dithered frames per epoch, the target position on CCD chip1, and constraints on airmass and atmospheric conditions, is very similar to the one of our exoplanet search survey \citep{Sahlmann:2014aa} and we reduced the data with the methods developed for that purpose \citep{Lazorenko:2009ph,Lazorenko:2014aa}. A particularity of this program is the choice of the OIII-6000 interference filter instead of the $I$-Bessel filter, which was necessary to decrease the contrast between the comparably bright \mystar\ and field stars.

\begin{figure} 
\centering
\includegraphics[width=\linewidth]{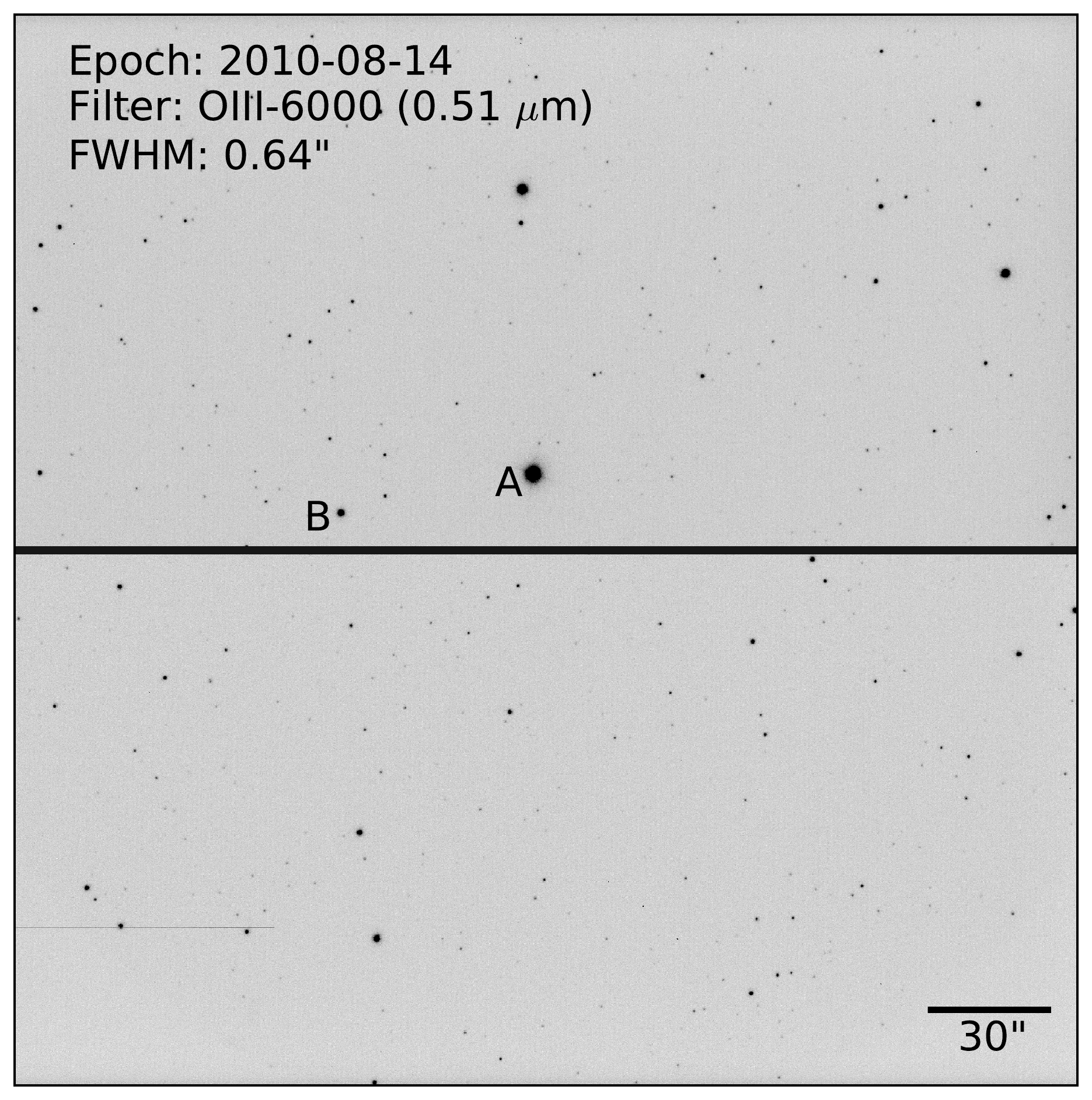}
\caption{FORS2 image of \mystar. The A and B components of GJ\,676 are labelled. The image size corresponds to the entire $4\arcmin\times4\arcmin$ field of view imaged onto two chips, with North up and East left.} 
\label{fig:image}
\end{figure}

Table \ref{tab:obs} summarises the data we obtained and lists the epoch number, the mean date of the epoch exposures, the average airmass and the average FWHM measured for star images. There are 8 epochs spanning 865 d, and every epoch consists of 49 to 77 usable individual exposures ($N_\mathrm{f}$) taken over $\Delta t=0.8$ h on average, resulting in a total of 517 exposures. Figure \ref{fig:image} shows an example image. We used 272 reference stars located within a radius of 2\farcm1 of \mystar\ ($I \simeq 8.6$ mag, $V \simeq 9.6$ mag) to measure the motion of the star relative to the background field. The majority of stars are faint and span a magnitude range of $I \sim 16 - 19$. Only four stars (including \mystarB) are relatively bright and 2--5 mag fainter than \mystar. 

The model for the astrometric reduction takes into account effects of various origin: instrumental (optical distortion, relative motion of the CCD chips), atmospheric (random image motion, differential chromatic refraction), and astrophysical, e.g.\ the displacement of reference stars due to proper motion and parallax. Optical distortion, in particular, is modeled by fitting the deformation  of  the reference star positions between frames with  basic functions that are polynomials in $x$, $y$ of powers $0,1, \ldots, k/2-1$. The even integer $k$ is  called the mode of the astrometric reduction and varies in the range of $6 \ldots 12$ which corresponds to polynomials of the power $2 \ldots 5$. The model is adjusted simultaneously to all available measurements and has thousands of free parameters, e.g.\ indexes of differential chromatic refraction and parallaxes of every star. The target object is not used for these reduction steps. Using the derived model, we then corrected the measured photocentre positions of the target object \mystar\ and extracted its astrometric parameters as described in Sect.\,{\ref{sec:astroparam}}. More details on the reduction principles can be found in \cite{Lazorenko:2006qf} and \cite{Lazorenko:2009ph, Lazorenko:2011lr, Lazorenko:2014aa}. Eventually, we obtained relative astrometric measurements of \mystar\ with an average per-epoch precision of 0.43 mas. {The epoch astrometry is given in Table \ref{tab:epastr}.}

 \mystar\ is a relatively bright target  and the exposure times have to be short to avoid saturation, reducing the number of measurable reference stars.  Consequently, the nominal astrometric errors compare unfavourably with the precisions obtained for the ultracool dwarfs that we survey for orbiting planets \citep{Sahlmann:2014aa}, mostly due to a lack of bright reference stars which resulted in higher reference-frame noise. One  bright reference star is  \mystarB. To slightly improve precision, we  set  the parallax of \mystarB\ to be equal to that  for \mystar, noting that the difference of parallaxes in this binary system is small. Because  the precision of the parallax measurement corresponds to $\approx$0.1\,pc in distance, the uncertainty of the distance measurement between these two stars is $\sim$20\,000 AU. This greatly exceeds the expected value for the relative binary separation, which can be estimated from the sky-projected separation between the two stars. For the measured angular separation of  $\sim$47\arcsec at the distance of $\sim$16.7\,pc, the expected value is 800 AU, thus unresolved with our astrometry. 

\begin{table}
\caption{FORS2 data used in the astrometric analysis.}
\centering
\begin{tabular}{rcrrrrcc}
\hline
\hline
No & Mean date & $N_\mathrm{f}$ & $\Delta t$ & Air- & FWHM \\
 & (UT) &  & (h) & mass& (\arcsec) \\
\hline
1 & {2010-04-10T07:21:30} & 71 & 0.77 & 1.19 & 0.67 \\
2 & {2010-05-11T06:31:57} & 67 & 0.78 & 1.13 & 0.62 \\
3 & {2010-07-09T02:40:03} & 49 & 0.76 & 1.13 & 0.75 \\
4 & {2010-08-14T00:08:57} & 77 & 0.76 & 1.13 & 0.64 \\
5 & {2011-04-11T09:10:05} & 64 & 0.75 & 1.12 & 0.67 \\
6 & {2011-06-09T04:00:17} & 54 & 0.65 & 1.15 & 0.74 \\
7 & {2012-07-23T03:36:52} & 67 & 0.80 & 1.18 & 0.66 \\
8 & {2012-08-22T00:38:43} & 68 & 0.79 & 1.13 & 0.70 \\
\hline
\end{tabular}
\label{tab:obs}
\end{table}

\subsection{HARPS radial velocities}
\cite{Forveille:2011lr} used 69 high-precision radial-velocity measurements of \mystar\ obtained with the HARPS instrument \citep{Mayor:2003cs} between 2006 and 2010 to discover planet `b`. We have since continued to observe \mystar\ regularly with HARPS and collected 60 additional measurements between 2010 and 2016, bringing the total number of radial-velocity datapoints to 129. Some of these new observations were taken as part of the volume limited survey \citep{Lo-Curto:2010fk} and of the follow-up on long-period planets \citep{Moutou:2015aa}. 

To extract radial velocities (RV), we constructed a high signal-to-noise spectrum by combining all \mystar\ spectra and computed the RV at each epoch with a chi-square minimisation relative to that master spectrum as described by \cite{Astudillo-Defru:2015aa}. Only for the two most recent measurements, we used the RV values given by the ESO standard instrument pipeline, which uses the cross-correlation function. The derived velocities used in this study are listed in Table \ref{tab:rv}. This is because those data were obtained after the HARPS upgrade, in which new octagonal fibres were installed \citep{Lo-Curto:2015aa}, which changed the line-spread function significantly and many more observations will be needed to generate a second master spectrum for observations taken after the HARPS upgrade. We thus treated the last two measurements as if they were taken with a different instrument and allowed for an offset in the model.

On 2016-03-05 (BJD 57452.837858) we also obtained one HARPS observation of \mystarB\ and measured its RV to $-39.3960 \pm 0.0034$\,\kmps, which was obtained with the standard pipeline and may include a small ($\lesssim1.5$\,\mps) zero-point offset from the measurements for \mystar.
  
\section{Analysis of astrometric data}  
\mystar\ (HIP\,85647) was observed 79 times over 1110 days by \emph{Hipparcos} \citep{ESA:1997vn} with a median astrometric uncertainty of $\sigma_{\Lambda}=6.1$\,mas \citep{:2007kx}, hence covering the orbit of \mystar\,b 1.05 times. Using the methods of combining the radial-velocity orbital parameters with the \emph{Hipparcos} astrometry described in \cite{Sahlmann:2011fk, Sahlmann:2011lr}, we found no orbital signature in the \emph{Hipparcos} astrometry data: both the permutation test and the F-test yield astrometric orbit significances below 1-$\sigma$. However, we can use the \emph{Hipparcos} observations to set an upper limit to the companion mass by determining the minimum detectable astrometric signal $a_\mathrm{min}$ of the individual target. When the data cover at least one complete orbit, \cite{Sahlmann:2011fk, Sahlmann:2011lr} showed that an astrometric signal-to-noise of $\mathrm{S/N} \gtrsim 6-7$ is required to obtain a detection at the 3-$\sigma$ level, where $\mathrm{S/N}=a\, \sqrt{N_\mathrm{Hip}} / \sigma_{\Lambda}$ and $a$ is the semi-major axis of the detected barycentric stellar orbit. Using a conservative S/N-limit of 8, we derive the upper companion mass limit $M_{2,\mathrm{max}}$ as the companion mass which introduces the astrometric signal $a_\mathrm{min} = 8 \sigma_{\Lambda} / \sqrt{N_\mathrm{Hip}} (1-e^2)$, where the factor $1-e^2$ accounts for the most unfavourable case of $i=90\degr$ and $\omega=90\degr$ in which the astrometric signal is given by the semi-minor axis of the orbit. Using this criterion and a primary mass of $M_1 = 0.71\,M_\sun$ \citep{Forveille:2011lr}, we set an upper limit of $44\,M_\mathrm{J}$ to the mass of \mystar\,b, i.e.\ this companion must be a substellar object.

\subsection{Analysis of FORS2 astrometry}\label{sec:astroparam}
We first analysed the FORS2 astrometry using the standard seven-parameter model without orbital motion as described in, e.g., Sect.\ 4.1 of \citet{Sahlmann:2014aa}. We obtained preliminary astrometric parameters of \mystar\ by obtaining the least-squares solution of Eq.\ (\ref{eq:axmodel}) for the photocentre positions determined from the FORS2 images. When neglecting orbital motion, this is a linear model and the solution was obtained using matrix-inversion, taking into account the measurement uncertainties and covariances. 

\begin{equation}\label{eq:axmodel}
\begin{array}{ll@{\hspace{1mm}}l@{\hspace{1mm}}l}
\!\!\alpha^{\star}_{m} =\!\!\!\!\!\!& \Delta \alpha^{\star}_0 + \mu_{\alpha^\star} \, t_m + \varpi \, \Pi_{\alpha,m} &- \rho\, f_{1,x,m}-  d \,f_{2,x,m} &+ (B \, X_m + G \, Y_m)\\
\!\delta_{m} = \!\!\!\!\!\!& \underbrace{\Delta \delta_0 + \,\mu_\delta      \,  \;                      t_m \;+ \varpi \, \Pi_{\delta,m}}_{\mbox{Standard model} }  &\underbrace{+ \rho \,f_{1,y,m}+  d \,f_{2,y,m}}_{\mbox{Refraction}} &\underbrace{+ (A \, X_m + F \, Y_m)}_{\mbox{Orbital motion}}.
\end{array}
\end{equation}

\begin{figure}[h!] 
\centering
\includegraphics[width=\linewidth]{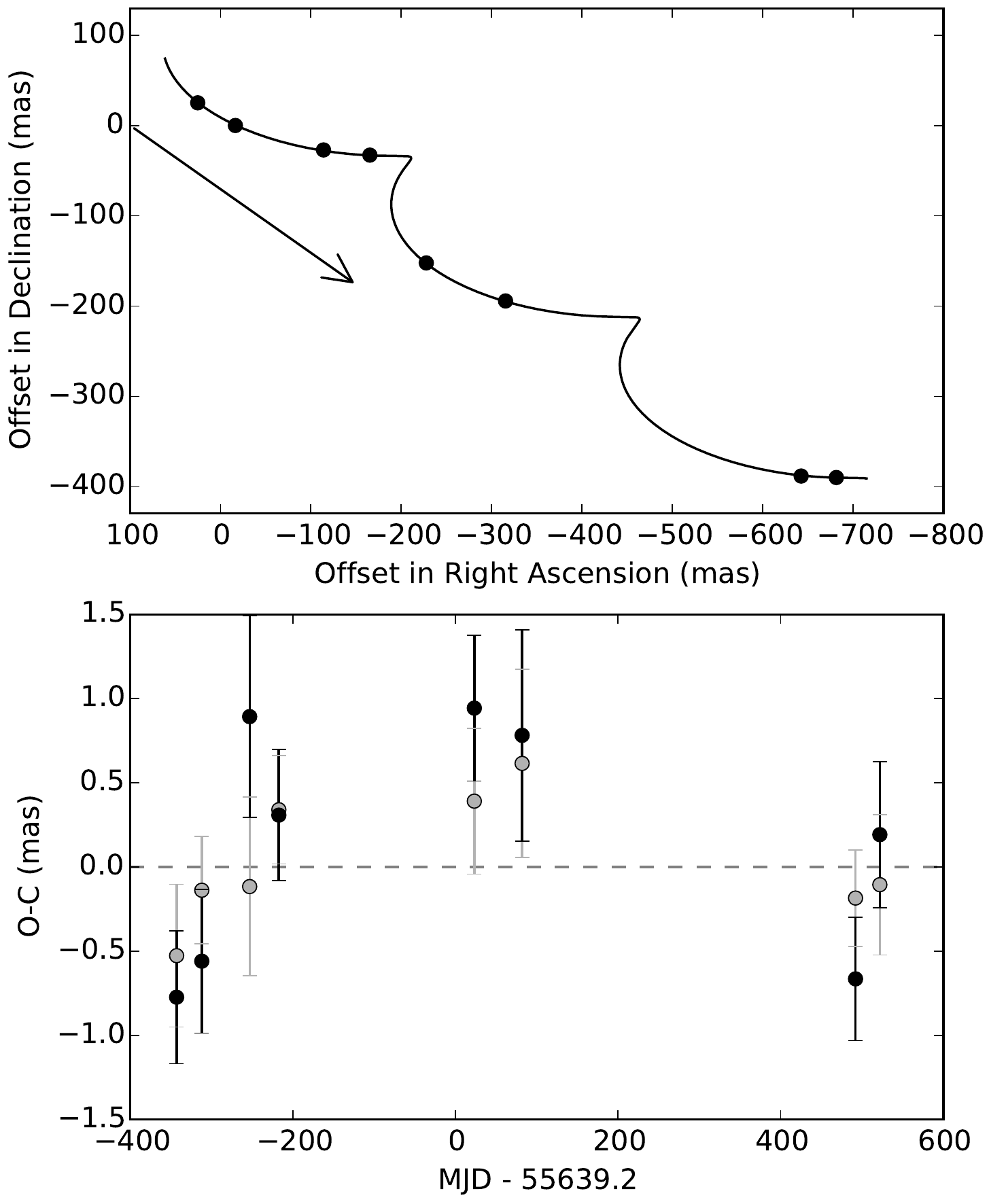}
\caption{\emph{Top:} The sky-projected motion of \mystar\ measured with FORS2. Epoch measurements are shown with black circles and the best-fit model is shown by the curve. The arrow indicates the proper motion per year. \emph{Bottom:} Epoch residuals in RA (grey symbols) and Dec (black symbols) of the seven-parameter fit as a function of time.}\label{fig:baryfit}
\end{figure}

The parameters are given in Table \ref{tab:ppm}, and Fig.\ \ref{fig:baryfit} illustrates the results. In Table \ref{tab:ppm} we also compare these results to the final adopted solution (see next Section) and to the \emph{Hipparcos} catalogue values, which shows that FORS2 astrometry yields proper motions and a parallax that are compatible with \emph{Hipparcos}. 

As can be seen in the bottom panel of Fig.\ \ref{fig:baryfit} the standard model does not fit the data well. The residual r.m.s. dispersion is 0.54 mas, which is larger than the average epoch precision of 0.43 mas and corresponds to a reduced $\chi^2_\mathrm{epoch}=2.6$. More importantly, the curved shape of the residuals hints towards a systematic effect rather than random noise. 

To investigate whether the excess correlated signal is associated to the stellar reflex motion caused by planet `b`, we performed a combined, yet sequential, analysis of radial-velocity and astrometric data, which we adapted from \cite{Sahlmann:2011fk}: Using the spectroscopic orbital parameters of \cite{Forveille:2011lr}, we fitted the FORS2 epoch astrometry that was corrected for DCR with a seven-parameter model, where the free parameters are the inclination $i$, the longitude of the ascending node $\Omega$,  the parallax $\varpi$, and offsets to the coordinates ($\Delta \alpha^{\star}$, $\Delta \delta$) and proper motions ($\Delta \mu_{\alpha^\star}$, $\Delta \mu_{\delta}$). A two-dimensional grid in $i$ and $\Omega$ was searched for its global $\chi^2$-minimum with a standard nonlinear minimisation procedure. The statistical significance of the derived astrometric orbit was determined with a permutation test employing 1000 pseudo orbits. Uncertainties in the solution parameters were derived by Monte Carlo simulations that include propagation of RV parameter uncertainties. This method has proven to be reliable in detecting orbital signatures in the \emph{Hipparcos} Intermediate Astrometric Data \citep[e.g.][]{Diaz:2012fk, Sahlmann:2013fk3, Wilson:2016aa}.

This analysis yielded an orbit significance of 99.8 \% corresponding to better than 3-$\sigma$ on the basis of the permutation test. Therefore the astrometric orbit is clearly detected with the FORS2 astrometry. The preliminary parameters derived with this method are an orbital inclination of $42\pm9\degr$, corresponding to a planet mass of $7.3 \pm 1.3\,M_\mathrm{J}$, and an ascending node of $\Omega_\mathrm{seq}=200\pm11\degr$. The residual r.m.s.\ of the best solution is 0.28 mas {(corresponding to a reduced $\chi^2$ of 0.61)} and significantly smaller than when employing the model without orbital motion. {We repeated the same analysis with the updated orbital parameters of planet `b` obtained with new HARPS RV data (see Sect.\ \ref{sec:RVanalysis}), which agrees with the \cite{Forveille:2011lr} solution within the uncertainties, and obtained essentially the same results.} To derive more accurate model parameters and uncertainties, we performed a joint analysis of radial velocity and astrometry data, which is presented in Sect.\ \ref{sec:mcmc}.

\begin{table*}
\caption{Astrometric parameters of \mystar.}
\centering
\begin{tabular}{ccr r r r r}
\hline
\hline
Par. & Unit & Standard model\tablefootmark{a} & Orbit model\tablefootmark{b} & {Orbit model\tablefootmark{c}} & HIP & HIP2 \\
& & (Linear fit) & (MCMC) & {(Sequential fit)} & (1) & (2) \\
\hline
$\Delta \alpha^\star_0$ & (mas) & $-265.13 \pm 0.2$ &$-264.1\pm0.2$ & N/A  & N/A& N/A        \\[2pt] 
$\Delta \delta_0      $  & (mas) & $-126.1 \pm 0.6$     &$-125.3\pm0.5$& N/A  &N/A &   N/A     \\ [2pt] 
$\varpi              $  & (mas) & $59.7 \pm 0.3$ &$59.3\pm0.3$ & $59.3\pm0.2$  &$ 61.98\pm1.81  $&$ 60.79 \pm	1.62  $ \\ [2pt] 
$\mu_{\alpha^\star}$  & (mas yr$^{-1}$) & $-252.9 \pm 0.3$\tablefootmark{d}  &$-253.4\pm0.4$\tablefootmark{d}& $-253.4\pm0.2$\tablefootmark{d} &$ -259.23 \pm1.46  $&$  -260.02 \pm	1.34 $ \\ [2pt] 
$\mu_{\delta}$         & (mas yr$^{-1}$) & $-178.2 \pm 0.2$\tablefootmark{d} &$-177.9\pm0.2$\tablefootmark{d} & $-177.9\pm0.2$\tablefootmark{d} &$ -185.69\pm0.92  $&$ -184.29 \pm	0.82   $ \\ [2pt] 
$\rho$                  & (mas) & $-31 \pm 6$  &  $-29\pm6$  & N/A&       N/A&       N/A     \\ [2pt] 
$d$                  & (mas) & $22 \pm 5$   &  $20\pm5$  & N/A&   N/A   &       N/A      \\ 
\hline
\end{tabular}
\label{tab:ppm}
\tablefoot{\tablefoottext{a}{Seven-parameter model without orbital motion. Standard uncertainties were computed from the parameter variances that correspond to the diagonal of the problem's inverse matrix and rescaled to take into account the residual dispersion.} \tablefoottext{b}{Adopted solution (see Sect. \ref{sec:mcmc}).} \tablefoottext{c}{{See Sect. \ref{sec:astroparam}.}} \tablefoottext{d}{Relative proper motion that cannot directly be compared to the \emph{Hipparcos} absolute proper motions.}\\\textbf{References}: (1) \cite{ESA:1997vn}; (2) \cite{:2007kx}.}
\end{table*}

\subsection{Parallax correction}
Because the astrometric reference stars are not located at infinity, a correction has usually to be applied to the relative parallax to convert it to absolute parallax that allows us to determine the distance to the system. As in \cite{Sahlmann:2014aa}, we used the Galaxy model of \cite{Robin:2003fk} to obtain a large sample of pseudo-stars in the region around \mystar. The comparison between the model parallaxes and the measured relative parallaxes of stars covering the same magnitude range yields an average offset, which is the parallax correction $\Delta \varpi_\mathrm{galax}$. The absolute parallax $\varpi_\mathrm{abs} = \varpi - \Delta \varpi_\mathrm{galax} $ is larger than the relative parallax because the reference stars absorb a small portion of the parallactic motion, i.e.\ the parallax correction has to be negative.

Using $N_\mathrm{s}\!=\!142$ reference stars, we obtained a parallax correction of $\Delta \varpi_\mathrm{galax}\!=\!+0.12 \pm 0.24$\,mas for \mystar. Because the reference stars are much fainter than \mystar, their parallaxes have large uncertainties, which translated into a large uncertainty of the parallax correction. The correction is smaller than its uncertainty, i.e.\ it is compatible with zero, and it has a positive value which is not allowed by definition. Therefore we did not apply the correction to the parallax of \mystar\ and we set  $\varpi_\mathrm{abs} = \varpi$.

In principle, a similar procedure should be applied to correct from relative to absolute proper motion. We refrain from doing so, because proper motions are not critical parameters in the following analyses and their corrections will be small. In the future, the results of ESA's \emph{Gaia} mission will make it possible to determine model-independent parallax and proper motion corrections, because \emph{Gaia} will obtain accurate astrometry for many of the reference stars and for \mystar\ and \mystarB\ themselves.

\begin{table}[h!]
\caption{{Results of the MCMC analysis of the HARPS radial velocities obtained with a four-Keplerian model. For planet `c`, we fitted a circular orbit, i.e.\ $e_\mathrm{c}=0$ and $\omega_\mathrm{c}=0$.} }
\centering
\begin{tabular}{ccrr}
\hline\hline
Parameter & Unit & This work &  \citetalias{Anglada-Escude:2012aa} \\
\hline
$\gamma$ & (\hbox{m s$^{-1}$}) & $-39038.0^{+1.1}_{-1.1}$ & N/A \\[3pt]
$\Delta \gamma_0$ & (\hbox{m s$^{-1}$}) & $2.0^{+1.3}_{-1.3}$  & N/A \\[3pt]

\multicolumn{4}{c}{Planet `b`}\\
$P_\mathrm{b}$ & (day) & $1051.1^{+0.5}_{-0.5}$ & $1050.3^{+1.2}_{-1.2}$ \\[3pt]
$e_\mathrm{b}$ &  & $0.323^{+0.002}_{-0.002}$ & $0.328^{+0.004}_{-0.004}$ \\[3pt]
$M_\mathrm{b}\sin{i}$ & ($M_\mathrm{J}$) & $4.713^{+0.009}_{-0.009}$ & $4.950^{+0.310}_{-0.310}$ \\[3pt]
$\omega_\mathrm{b}$ & ($^\circ$) & $86.9^{+0.4}_{-0.4}$ & $87.4^{+0.7}_{-0.7}$ \\[3pt]
$T_{\mathrm{b},\mathrm{P}}$ & (day) & $55409.3^{+0.8}_{-0.8}$ & N/A \\[3pt]
$K_{1,\mathrm{b}}$ & (\hbox{m s$^{-1}$}) & $124.5^{+0.3}_{-0.3}$ & $117.42\pm0.42$ \\[3pt]

\hline
\multicolumn{4}{c}{Planet `c`}\\
$\log{P_\mathrm{c}}$ & (day) & $3.87^{+0.01}_{-0.01}$ & 3.64 \\[3pt]
$P_\mathrm{c}$ & (day) & $7462.9^{+105.4}_{-101.4}$ & 4400\\[3pt]
$M_\mathrm{c}\sin{i}$ & ($M_\mathrm{J}$) & $6.9^{+0.1}_{-0.1}$ & 3.0 \\[3pt]
$T_{\mathrm{c},\mathrm{P}}$ & (day) & $50404.9^{+63.5}_{-65.6}$ & N/A \\[3pt]
$K_{1,\mathrm{c}}$ & (\hbox{m s$^{-1}$}) & $90.0^{+1.2}_{-1.2}$ & 41 \\[3pt]

\hline
\multicolumn{4}{c}{Planet `d`}\\
$P_\mathrm{d}$ & (day) & $3.6005^{+0.0002}_{-0.0002}$ & $3.6000^{+0.0008}_{-0.0008}$ \\[3pt]
$e_\mathrm{d}$ &  & $0.262^{+0.090}_{-0.101}$ & $0.150^{+0.090}_{-0.090}$ \\[3pt]
$M_\mathrm{d}\sin{i}$ & ($M_\mathrm{J}$) & $0.014^{+0.001}_{-0.001}$ & $0.014^{+0.002}_{-0.002}$ \\[3pt]
$M_d\sin{i}$ & ($M_\oplus$) & $4.4^{+0.3}_{-0.3}$ & $4.4\pm0.7$ \\[3pt]
$\omega_\mathrm{d}$ & ($^\circ$) & $-48.7^{+13.8}_{-16.3}$ & $315.1^{+108.9}_{-108.9}$ \\[3pt]
$T_{\mathrm{d},\mathrm{P}}$ & (day) & $55498.7^{+0.1}_{-0.1}$ & N/A \\[3pt]
$K_{1,\mathrm{d}}$ & (\hbox{m s$^{-1}$}) & $2.4^{+0.2}_{-0.2}$ &  $2.30\pm0.32$\\[3pt]

\hline
\multicolumn{4}{c}{Signal/Planet `e`}\\
$P_\mathrm{e}$ & (day) & $35.39^{+0.03}_{-0.04}$ & $35.37^{+0.07}_{-0.07}$ \\[3pt]
$e_\mathrm{e}$ & () & $0.125^{+0.119}_{-0.087}$ & $0.240^{+0.120}_{-0.120}$ \\[3pt]
$M_\mathrm{e}\sin{i}$ & ($M_\mathrm{J}$) & $0.025^{+0.002}_{-0.002}$ & $0.036^{+0.005}_{-0.005}$ \\[3pt]
$M_e\sin{i}$ & ($M_\oplus$) & $8.1^{+0.7}_{-0.7}$ & $11.5\pm1.5$  \\[3pt]
$\omega_\mathrm{e}$ & ($^\circ$) & $331.7^{+19.7}_{-57.9}$ & $332.3^{+126.1}_{-126.1}$ \\[3pt]
$T_{\mathrm{e},\mathrm{P}}$ & (day) & $55509.2^{+1.9}_{-5.6}$ & N/A \\[3pt]
$K_{1,\mathrm{e}}$ & (\hbox{m s$^{-1}$}) & $2.0^{+0.2}_{-0.2}$ &  $2.62 \pm0.32$\\[3pt]

\hline
\end{tabular}
\label{tab:solutionRV}
\end{table}

\section{Analysis of radial velocities}\label{sec:RVanalysis}
\begin{figure}[h!] 
\centering
\includegraphics[width=\linewidth]{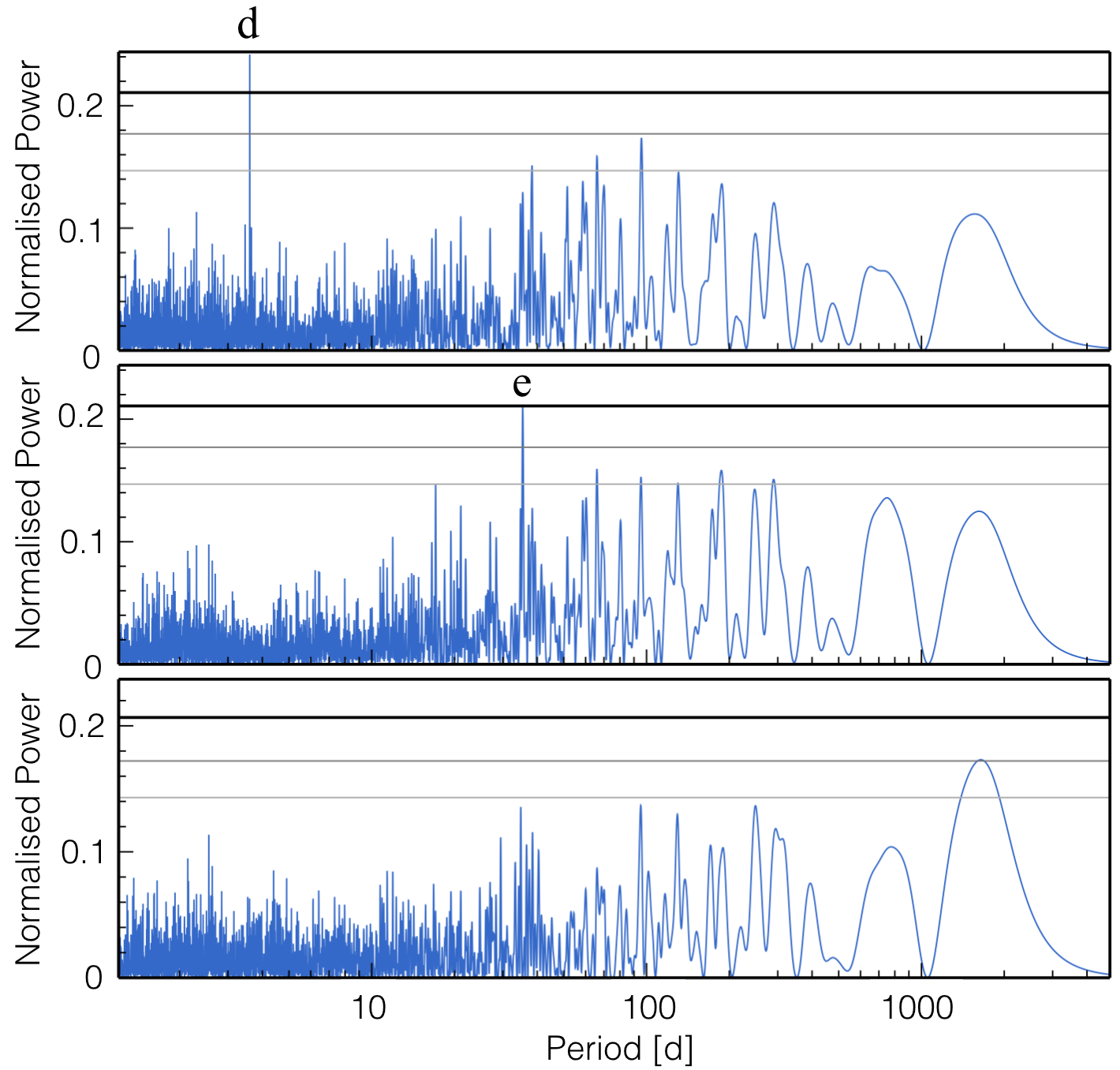}
\caption{Periodograms of the RV residuals after subtracting planets `b`+`c` (top panel), planets `b`+`c`+`d` (middle panel), and planets `b`+`c`+`d`+`e` (bottom panel). See text for discussion.}
\label{fig:periodograms}
\end{figure}

We analysed the radial velocities by using the {generalised Lomb-Scargle periodogram  \citep{Zechmeister:2009aa}} iteratively. We started by modelling the raw time series with one Keplerian (for planet `b`) plus a cubic drift (for planet `c`) and we computed the periodogram of the residuals of the best fit (top panel of Fig.\ \ref{fig:periodograms}). A powerful peak is seen around a period  of 3.6 days. To give a false-alarm probability (FAP), we assumed that the residuals are caused by random noise and we generated virtual data sets by swapping the radial-velocity values randomly while retaining their dates. For every virtual set we computed a new periodogram and measured the power of the highest peak. In this way we obtained the statistical distribution of power maxima that is expected from a timeseries that contains solely noise. From that distribution, the power values corresponding to a FAP of 1, 10, and 50 \% are the power values found to be greater than 99, 90, and 50 \% of the distribution, respectively. In Fig.\ \ref{fig:periodograms} those power levels are drawn with gray, dark-gray, and black lines, respectively, and the peak corresponding to planet `d` is recover with a FAP $\ll$1 \%. We next included an additional planet and applied a model composed of 2 Keplerians plus a cubic drift. The most prominent peak now has a period of $\sim$36 days and a FAP marginally below 1 \%. After iterating once more with yet an additional planet, no significant-power periodicity is seen in the final residuals. 
The most powerful peak in the last residual periodogram is located at $\sim$1600 days and has a FAP of $\sim$10 \%. If we were to interpret this as the signature of a yet undiscovered planet `f` and modelled it accordingly, the corresponding mass is $\sim$35\,$\,M_\oplus$, {i.e}.\  about two Neptune masses. 

\begin{figure*}
\centering
\includegraphics[width=\linewidth]{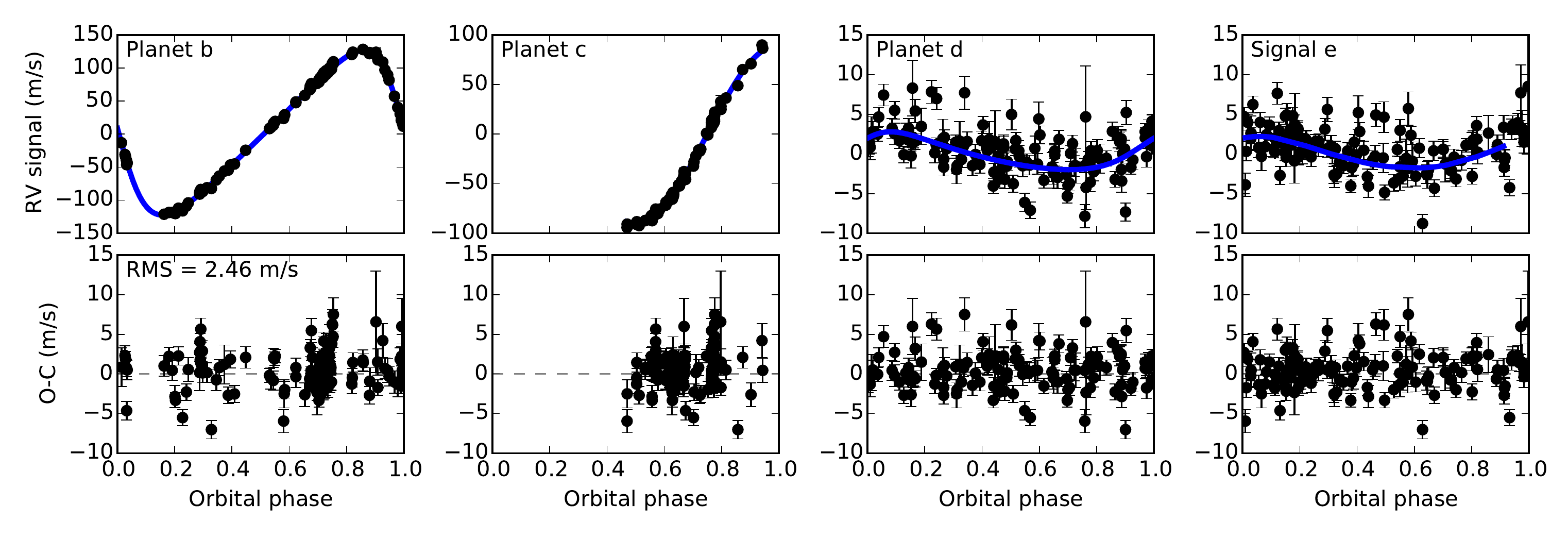}
\caption{{Radial-velocity signatures of the four Keplerians fitted to the \mystar\ measurements (top row) and the residuals as a function of respective orbital phase. Black circles with uncertainties show the HARPS measurements and the blue curves indicate the best-fit model.}}
\label{fig:RVsol}
\end{figure*}

{To derive the model parameters corresponding to the planetary signals, we performed a Markov Chain Monte Carlo (MCMC) analysis of the HARPS radial velocities. The model incorporated three Keplerians (planets `b`$+$`d` and signal `e`), one circular Keplerian for planet `c` (we found that both a circular Keplerian and a cubic drift models reproduce the signal equally well), the systemic velocity, and an offset to account for the HARPS upgrade (see Sect.\ \ref{sec:mcmc}), for a total of 20 free parameters. The parameter values are reported in Table \ref{tab:solutionRV} and compared to the solution presented by \citetalias{Anglada-Escude:2012aa}. Figure \ref{fig:RVsol} shows the four Keplerian curves and the data. Due to the larger number of measurements, our parameters for planets `b` and `d` are generally more precise but in good agreement with the \citetalias{Anglada-Escude:2012aa} parameters. For planet `c` we could derive the first, yet preliminary, good constraints on period and minimum mass, which are further discussed in Sect.\ \ref{sec:mcmc}. In comparison to \citetalias{Anglada-Escude:2012aa} for the signal attributed to planet `e`, we find a lower eccentricity and a smaller signal amplitude, hence a smaller minimum planet mass of $8.1\pm0.7$\,$M_\oplus$.}

\begin{figure}[h!] 
\centering
\includegraphics[width=0.9\linewidth]{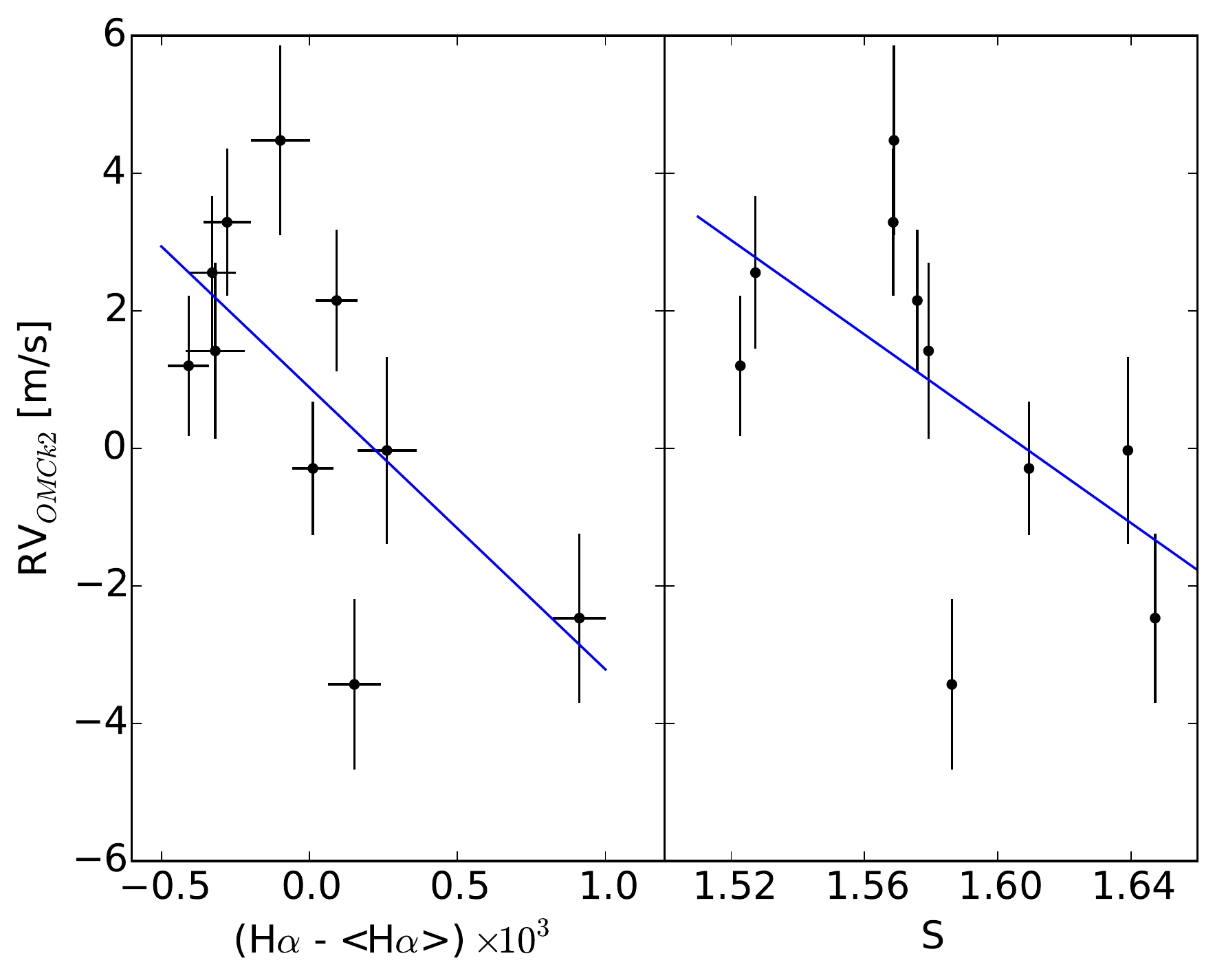}
\caption{RV residuals after subtracting `b`+`c`+`d` in the range MJD = 54660 -- 54690 as a function of the H$\alpha$ and S-index activity indicators. The solid lines show the best linear fits, which indicate negative correlations.} 
\label{fig:correl}
\end{figure}

\begin{figure}[h!] 
\centering
\includegraphics[width=0.9\linewidth]{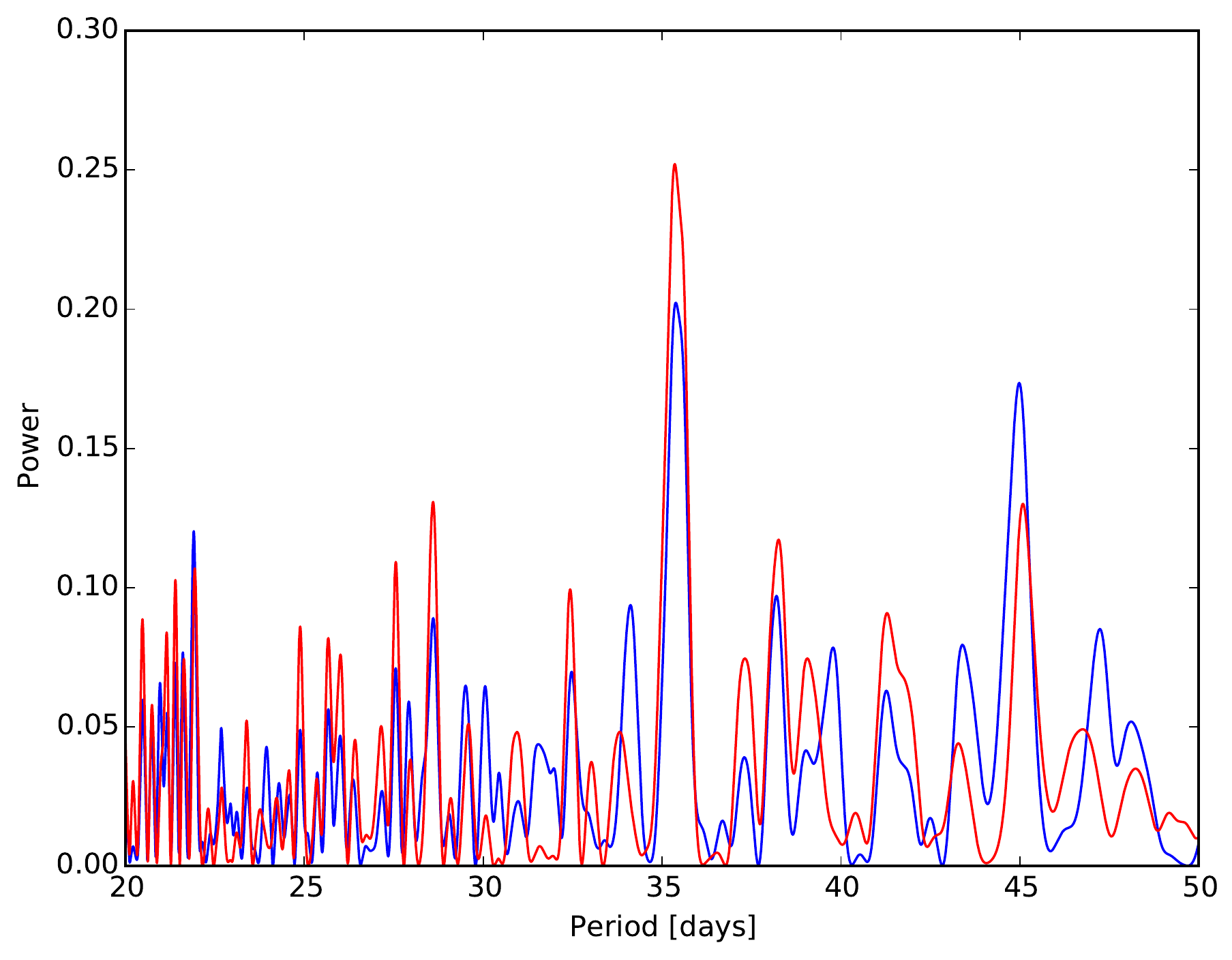}
\caption{Residual periodogram for model that includes planets `b`+`c`+`d`. The signal `e` is located at $\sim$36 d. The result with the original RV dataset is shown in red, whereas the blue curve was obtained after removing 10 measurements in the range MJD = 54660 -- 54690. Removing only those 10 points, which seem to correlate with activity indicators, decreases the power of signal `e` by as much as 20 \%.} 
\label{fig:correlations}
\end{figure}

{We} thus have recovered all signals previously reported by \cite{Forveille:2011lr} and \citetalias{Anglada-Escude:2012aa}. Before accepting the inner low-mass planets as genuine, however, we need to evaluate if the corresponding signals can alternatively be attributed to stellar activity. In particular, the period of planet  `e` ($\sim$36 d) is close to the stellar rotation period of $41.2\pm3.8$ d measured by \cite{Suarez-Mascareno:2015aa}. We also measured a chromospheric activity indicator $\log R'_{HK}=-4.599$ that is very close to the one of \object{GJ\,205} ($\log R'_{HK}=-4.596$), that has a known rotation period of 33.6 d \citep{Kiraga:2007aa}. The rotation period of \mystar\ is thus close to the orbital period of planet `e` and may induce a signal that is confused with the one of a planet. Furthermore, looking at the radial-velocity residuals when adjusting for all planets except for planet `e`, we noticed a strong correlation between the radial velocity and the activity index for epochs MJD 54660 -- 54690, see Fig.\ \ref{fig:correl}. This range includes only 10 radial-velocity points, but when removing just those 10 points from the original time series, the periodogram power of signal `e` decreases by as much as 20 \% as shown in Fig.\ \ref{fig:correlations}. We inspected several other activity indicators, e.g.\ the width of the cross-correlation function and its contrast, the S index, H$\alpha$ and Sodium indices, and were not able to identify the stellar rotation period. In summary, we remain cautious with the interpretation of signal `e` as being caused by a planet, because its period is close to the star's rotation period. Additional data and analyses are required to undoubtably establish the planetary nature of signal `e`. On the contrary, the period of planet `d` is sufficiently short compared to the stellar rotation period to be accepted as a planet.

\section{Joint analysis of radial velocities and astrometry}\label{sec:mcmc}
We applied an MCMC analysis to the individual radial-velocity data in Table \ref{tab:rv} and the astrometric measurements from our FORS2 observations. We used a global model with 21 free parameters that has five components:
\begin{itemize}
  \item \textbf{Radial velocity orbit of planet `b`:} There are 5 orbital parameters: the period $P$, eccentricity $e$, argument of periastron $\omega$, time of periastron passage $T_\mathrm{P}$, companion mass $M_\mathrm{b}$, and one offset $\gamma_0$ corresponding to the systemic velocity. 
  \item \textbf{Astrometric orbit of planet `b`:} The model comprises seven free parameters, of which five are shared with the radial velocity model ($P,e,\omega,T_P,M_\mathrm{b}$) and two are uniquely constrained by astrometry: the inclination $i$ and the ascending node $\Omega$. We also included two nuisance parameters $s_\alpha$ and $s_\delta$ for the astrometry in RA and Dec, respectively, to account for the off-diagonal terms in the covariance matrix of the FORS2 astrometry in individual frames \citep{Sahlmann:2013ab}. Because of the long period of planet `c`, its potential {non-linear} astrometric signature is much smaller than our uncertainties and {the linear part} will be absorbed by a bias in the proper motion values {(see Sect.\ \ref{sec:bias})}.
  \item \textbf{Radial velocity signature of planet `c`:} The orbital period of planet `c` is longer than the observation timespan (see Sect.\ \ref{sec:RVanalysis}). We modelled its signature with a circular Keplerian model ($e=0$, $\omega=0$) that has three parameters: orbital period $P_\mathrm{c}$, minimum mass $M_\mathrm{c} \sin{i}$, and time of {ascending node} $T_\mathrm{c,P}$.
  \item \textbf{Parallax and proper motion:} The standard astrometric model has five free parameters (position offsets $\Delta \alpha^\star_0$,$\Delta \delta_0 $, parallax $\varpi$, proper motions $\mu_{\alpha^\star}$,$\mu_\delta$) plus two parameters modelling differential chromatic refraction $\rho$ and $d$.
 \item \textbf{Radial velocity offset after HARPS upgrade:} Two RV measurement were taken with HARPS after the fibre upgrade \citep{Lo-Curto:2015aa}. To account for a potential instrumental offset, we included the free parameter $\Delta \gamma_0$ which is added to the RV data taken in the new HARPS configuration. With the help of 5 other stars with spectral types of M0--M4 observed with HARPS both before and after the upgrade, we found that the RV offset is compatible with zero with an uncertainty of 1.5 \mps.
 
\end{itemize}
We used the \texttt{emcee} package \citep{Foreman-Mackey:2013aa} to implement the MCMC and expressed the global model with the parameter vector $\teta$ composed of $P$, $e$, $\omega$, $T_\mathrm{P}$, $\gamma_0$, $\Delta \gamma_0$, $\log P_\mathrm{c}$, $M_\mathrm{c} \sin{i}$, $T_\mathrm{P,c}$, $\Delta \alpha^\star_0$, $\Delta \delta_0 $, $\varpi$, $\mu_{\alpha^\star}$, $\mu_\delta$, $\rho$, $d$, $s_\alpha$, $s_\delta$, $M_\mathrm{b} \sin{i}$, $M_\mathrm{b} \cos{i}$, and $\Omega$, where we chose the pair $M_\mathrm{b} \sin{i}$---$M_\mathrm{b} \cos{i}$ instead of $M_\mathrm{b}$---$i$ to mitigate the effect of the strong correlation that naturally exists between those parameters. The host star mass $M_1$ was kept constant. The logarithm of the likelihood as a function of the parameter vector $\teta$ was expressed as
\begin{equation}
\ln \mathcal{L}(\teta) = -0.5 \left( \chi^2_\mathrm{RV} + \chi^2_\mathrm{AX} \right) - \ln \mathcal{L}_0,
\label{eq.likelihood}
\end{equation}
where the subscripts RV and AX denote the radial-velocity and astrometric part of the model, respectively.
The radial velocity model $\mathcal{M}_\mathrm{RV}(\teta)$ implements the Keplerian equations for planets `b` and `c` and the corresponding $\chi^2$ is computed in the standard way
\begin{equation}
\chi^2_\mathrm{RV} = \sum_i \frac{ \left( \mathcal{M}_{i,\mathrm{RV}}(\teta) -  \mathcal{D}_{i,\mathrm{RV}} \right)^2 }{\sigma^2_{i,\mathrm{RV}}}
\label{eq:chi2rv}
\end{equation}
where $\mathcal{D}$ denotes measured quantities and $\sigma_i^2$ is the variance associated with the $i$-th measurement. For astrometry, the model $\mathcal{M}_\mathrm{AX}(\teta)$ implements Eq.\ (\ref{eq:axmodel}) and we account for the nuisance parameters $s_\alpha$ and $s_\delta$ when computing the total variance of an individual measurement and in the $\ln \mathcal{L}_0$ term

\begin{equation}
\chi^2_\mathrm{AX} = \sum_j \frac{ \left( \mathcal{M}_{j,\mathrm{AX}}(\teta) -  \mathcal{D}_{j,\mathrm{AX}} \right)^2 }{\sigma^2_{j,\mathrm{AX}} + s_{\alpha,\delta}^2}
\label{eq:chi2ax}
\end{equation}

\begin{equation}
\ln \mathcal{L}_0 = \sum_k \ln \left( \sqrt{2\pi} \sqrt{\sigma^2_{k,\mathrm{AX}} + s_{\alpha,\delta}^2} \right).
\label{eq:lnl0}
\end{equation}

For some parameters we applied uniform priors (see Table~\ref{tab:priors}) and imposed the following range limits: $0\degr \leqslant \Omega < 360\degr$, $M_\mathrm{b} \sin{i} > 0$, $\varpi > 0$, $\log{5000} < \log P_\mathrm{c} < \log{9000}$, $s_\alpha > 0$, and $s_\delta > 0$. For the RV offset $\Delta \gamma_0$, we applied a Gaussian prior centred on zero with a width of 1.5 \mps. For a more general discussion on combined modelling of radial-velocity and astrometric data see \cite{Wright:2009lr} and \cite{Anglada-Escude:2012vn}.

\begin{table}[h!]
\caption{List of priors.} 
\centering
\begin{tabular}{ccc}
\hline\hline
Parameter & Unit & Prior distribution \\
\hline
$\Omega$ 	&(\degr)	 		&$U(0; 360)$\\
$M_\mathrm{b} \sin{i}$ & ($M_\mathrm{J}$) &$U(0; \infty)$\\
$\varpi$ & (mas)&$U(0; \infty)$\\
$\log P_\mathrm{c}$ &(day) &$U(\log{5000}; \log{9000})$\\
$s_\alpha$& (mas)&$U(0; \infty)$\\
$s_\delta$ &(mas)&$U(0; \infty)$\\
$\Delta \gamma_0$ & (\mps) & $N(0; 1.5)$\\
\hline
\end{tabular}
\label{tab:priors}
\tablefoot{$U(x_{min};  x_{max})$: uniform distribution between $x_{min}$ and $x_{max}$. $N(\mu; \sigma)$: normal distribution with mean $\mu$ and standard deviation $\sigma$.}
\end{table}

We did not include the astrometric orbit terms of planets `d` and `e` in the model, because their signatures are of order 0.05 and 0.5 micro-arcsecond, respectively, (estimated for edge-on orbits with the parameters of \citetalias{Anglada-Escude:2012aa}), thus are negligible. Likewise, we did not include their radial-velocity terms (discussed in Sect.\ \ref{sec:RVanalysis}) because they have sufficiently small amplitudes and short periods that their omission does not affect the parameters of the large-amplitude and long-period signals of planets `b` and `c`.

Each of 160 walkers was initialised with a set of parameter values that was determined from the radial-velocity orbit parameters, from the standard astrometric fit in Sect.\ \ref{sec:astroparam}, and from the preliminary values for orbit inclination and planet mass derived from the sequential analysis in Sect.\ \ref{sec:astroparam}. Each walker was allowed to take 30\,000 steps, of which we discarded the first 25~\%. The solution was therefore derived from distributions with $3.6\cdot10^6$ samples. Table \ref{tab:solution} lists the adopted solution parameters determined as the median of the posterior distributions with 1-$\sigma$ uncertainties. 

\begin{table}
\caption{Solution derived from the MCMC. The actual model parameters are given in the text. For planet `c`, we fitted a circular orbit, i.e.\ $e_\mathrm{c}=0$ and $\omega_\mathrm{c}=0$. } 
\centering
\begin{tabular}{ccr}
\hline\hline
Parameter & Unit & Value \\
\hline
$\Delta \alpha^\star_0$ & (mas) & $-264.1^{+0.2}_{-0.2}$ \\[3pt]
$\Delta \delta_0 $ & (mas) & $-125.3^{+0.5}_{-0.5}$ \\[3pt]
$\varpi$ & (mas) & $59.3^{+0.3}_{-0.3}$ \\[3pt]
$\mu_{\alpha^\star}$ & (\hbox{mas yr$^{-1}$}) & $-253.4^{+0.4}_{-0.4}$ \\[3pt]
$\mu_\delta$ & (\hbox{mas yr$^{-1}$}) & $-177.9^{+0.2}_{-0.2}$ \\[3pt]
$\rho$ & (mas) & $-29^{+  6}_{-  6}$ \\[3pt]
$d$ & (mas) & $ 20^{+  5}_{-  5}$ \\[3pt]
$s_\alpha$ & (mas) & $0.3^{+0.3}_{-0.2}$ \\[3pt]
$s_\delta$ & (mas) & $0.4^{+0.4}_{-0.3}$ \\[3pt]
$\gamma_0$ & (\hbox{m s$^{-1}$}) & $-39038.3^{+1.0}_{-1.0}$ \\[3pt]
$\Delta \gamma_0$ & (\hbox{m s$^{-1}$}) & $0.5^{+1.2}_{-1.3}$ \\[3pt]
Dist. & (pc) & $16.9^{+0.1}_{-0.1}$ \\[3pt]
$T_\mathrm{Ref}$ &(MJD)& {55637.693209}\\[3pt]
\hline
\multicolumn{3}{c}{Planet `b`}\\
$P$ & (day) & $1052.1^{+0.4}_{-0.4}$ \\[3pt]
$e$ &  & $0.323^{+0.002}_{-0.002}$ \\[3pt]
$M_\mathrm{b}\sin{i}$ & ($M_\mathrm{J}$) & $4.733^{+0.011}_{-0.010}$ \\[3pt]
$M_\mathrm{b}\cos{i}$ & ($M_\mathrm{J}$) & $4.7^{+2.3}_{-2.6}$ \\[3pt]
$T_\mathrm{P}$ & (day) & $55410.4^{+0.8}_{-0.8}$ \\[3pt]
$\omega$ & ($^\circ$) & $87.4^{+0.4}_{-0.4}$ \\[3pt]
$\Omega$ & ($^\circ$) & $208^{+ 15}_{- 13}$ \\[3pt]
$i$ & ($\degr$) & $ 45^{+ 21}_{- 11}$ \\[3pt]
$a_1$ & (mas) & $1.0^{+0.3}_{-0.2}$ \\[3pt]
$a_{rel}$ & (mas) & $107.5^{+0.4}_{-0.4}$ \\[3pt]
$a_\mathrm{rel}$ & (AU) & $1.812^{+0.002}_{-0.001}$ \\[3pt]
$M_\mathrm{b}$ & ($M_\mathrm{J}$) & $6.7^{+1.8}_{-1.5}$ \\[3pt]
\hline
\multicolumn{3}{c}{Planet `c`}\\

$T_{c,P}$ & (day) & $50495.2^{+57.6}_{-60.2}$ \\[3pt]
$M_c\sin{i}$ & ($M_\mathrm{J}$) & $6.8^{+0.1}_{-0.1}$ \\[3pt]
$\log{P_c}$ & (day) & $3.87^{+0.01}_{-0.01}$ \\[3pt]
$P_\mathrm{c}$ & (day) & $7337^{+ 95}_{- 92}$ \\[3pt]
$a_\mathrm{rel,c}$ & (AU) & $6.6^{+0.1}_{-0.1}$ \\[3pt]
$K_{1,\mathrm{c}}$ & (\hbox{m s$^{-1}$}) & $88.7^{+1.1}_{-1.1}$ \\[3pt]

\hline
\end{tabular}
\label{tab:solution}
\end{table}

\subsection{Parallax and relative proper motions}
Our parallax determination agrees with the \emph{Hipparcos} parallax within the uncertainties, yet it is slightly smaller leading to a larger distance of $16.86\pm0.07$ pc to \mystar. This in turn would lead to a slightly higher mass estimate for the star, however, we assume a 10 \% uncertainty for the primary mass determination $M_1=0.71\,M_\sun$ (\citealt{Forveille:2011lr}, \citetalias{Anglada-Escude:2012aa}), which renders this adjustment insignificant. 

Our relative proper motions are discrepant from the absolute measurements derived by \emph{Hipparcos} \citep{:2007kx} at the 5-$\sigma$ -- 8-$\sigma$ level (cf.\ Table \ref{tab:ppm}). This can be explained by the intrinsically relative measurements accessible by FORS2, but also by the fact that \emph{Hipparcos} and FORS2 proper motions are biased differently by the orbital motion of planet `c`. The high precision of our relative proper motion measurement allowed us to perform a detailed study of the relative orbit of the wide binary, which is presented in Sect.\ \ref{sec:AB}.

\begin{figure}[h!]
\centering
\includegraphics[width=\linewidth]{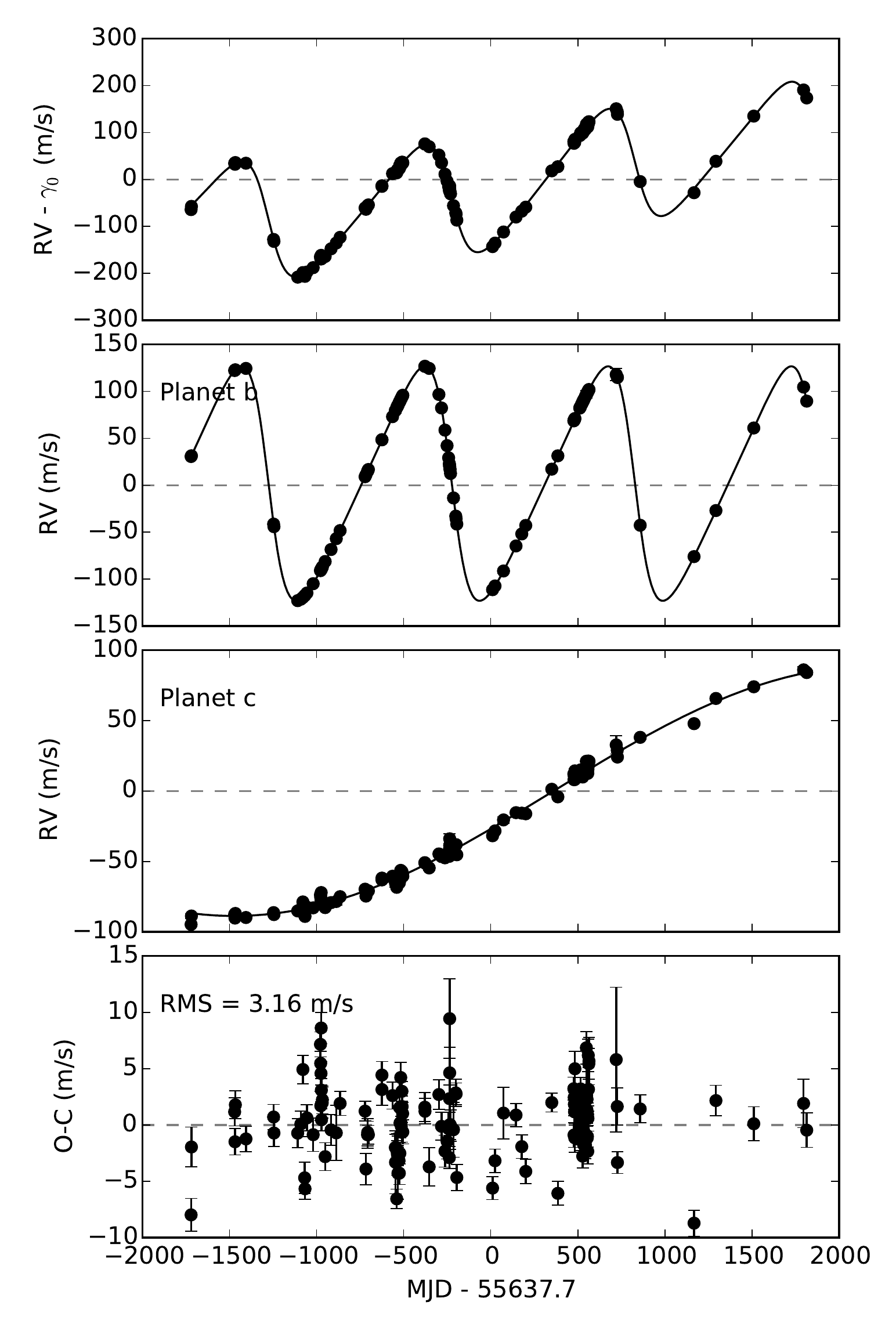}
\caption{Measured radial velocities of \mystar\ (top panel) and the residuals after subtracting the best-fit model indicated by the solid line (bottom panel). The two middle panels display the radial velocity signatures of planets `b` and `c`.}
\label{fig:rvomc}
\end{figure}

\subsection{Radial velocity orbits} 
The radial velocity data and the best-fit model are shown in Fig.~\ref{fig:rvomc}. The orbital parameters of planet `b` in Table \ref{tab:solution}  are compatible with the estimates of \citet{Forveille:2011lr} and \citetalias{Anglada-Escude:2012aa}, yet they are more precise because we have more data available. In particular, the minimum mass of planet `b` is $M_\mathrm{b}\sin{i} = 4.733 \pm 0.011\,M_\mathrm{J}$. The residual r.m.s.\ of our model that does not include any of the two inner planets `d` and `e` is 3.16 \mps. 

Although our RV data do not cover one full revolution of planet `c`, its orbital period and RV amplitude is relatively well constrained. Assuming a circular orbit, we find a period of $7337\pm95$ days ($\sim$20 years) with a semi-amplitude of $\sim$89 \mps, which corresponds to a minimum planet mass of $6.8\pm0.1$\,$M_\mathrm{J}${, where the uncertainty does not include the 10 \% uncertainty on the mass of the star}. The corresponding relative semimajor axis is $6.6\pm0.1$ AU, placing planet `c` between the orbital distances of Jupiter and Saturn in the Solar System. Until the eccentricity of planet c's orbit can be determined with additional measurements, these values should be considered preliminary.

\subsection{Astrometric orbit and the mass of \mystar\,b}
The MCMC chains converged towards stable solutions that produced quasi-Gaussian posterior distributions for most of the 21 free parameters, which indicated that the model is well-constrained and the astrometric orbit was detected with our FORS2 measurements. Figure \ref{fig:corner} displays the joint marginal distributions of all MCMC parameters and shows that correlations are generally weak, with the exception of the expected inter-dependencies between the periastron parameters $T_P$ and $\omega$ and the chromatic refraction parameters $\rho$ and $d$ that are anti-correlated by design.

\begin{figure}[h!]
\centering
\includegraphics[width=\linewidth]{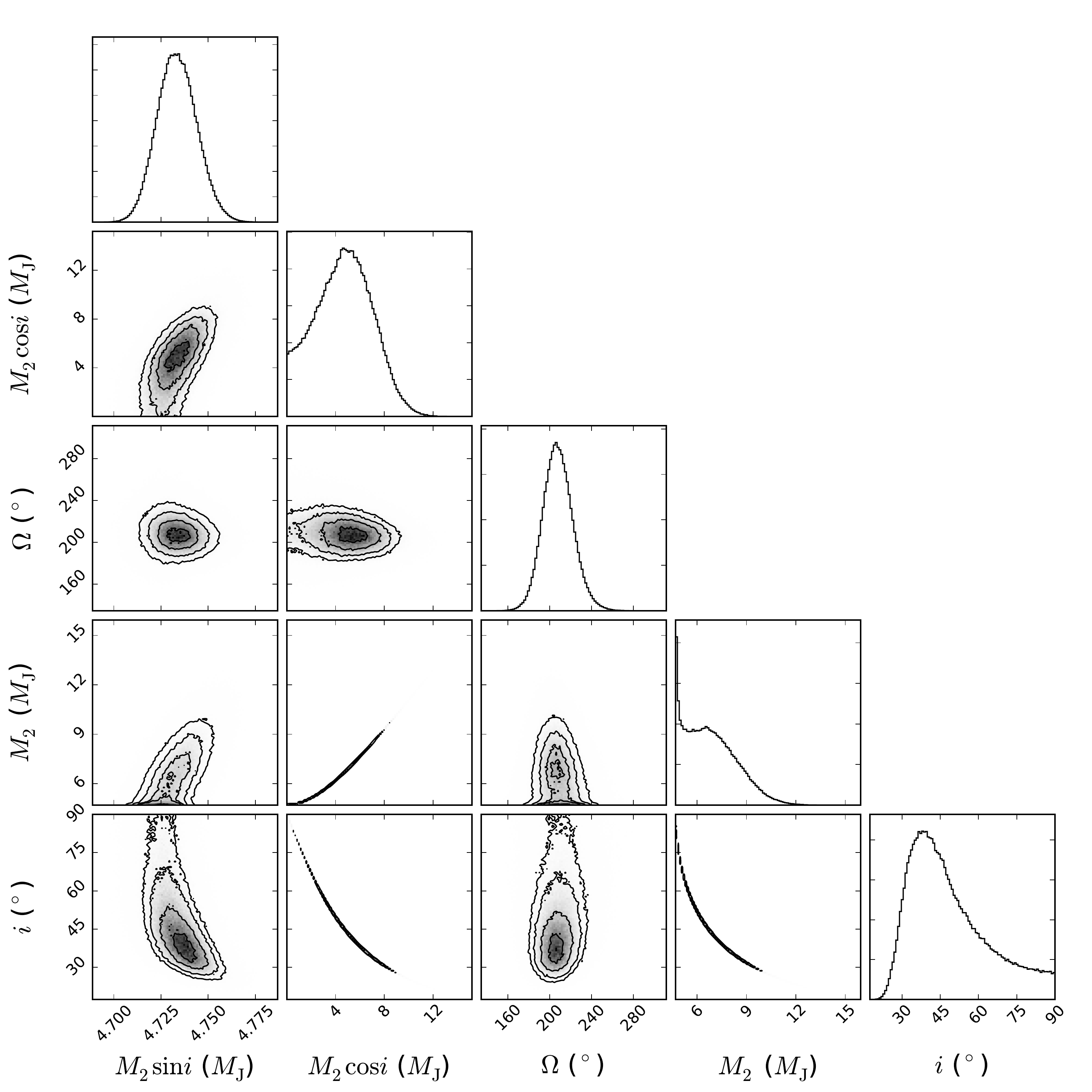}
\caption{The joint marginal distributions of the three free parameters probed by the MCMC that determine the orbital inclination, hence mass, of planet `b` and the sky orientation $\Omega$ of the orbit. The parameters $M_\mathrm{b}$ and $i$ are computed from $M_\mathrm{b}\sin{i}$ and $M_\mathrm{b}\cos{i}$.}
\label{fig:corner2}
\end{figure}

The joint marginal distributions of orbit parameters that are constrained by astrometry are shown in Fig.\ \ref{fig:corner2}. All three ($\Omega$, $M_\mathrm{b} \sin{i}$, $M_\mathrm{b} \cos{i}$) are well constrained and are weakly correlated. After conversion to the $M_\mathrm{b}$ and $i$ parameters, which are also displayed in Fig.\ \ref{fig:corner2}, the lower mass limit imposed by the radial velocities becomes apparent as a pile-up at $\sim$$4.7 \,M_\mathrm{J}$, but a peak of the mass distribution at $\sim$$6.7 \,M_\mathrm{J}$ can be clearly identified. 

\begin{figure*}
\sidecaption
\includegraphics[width=12cm]{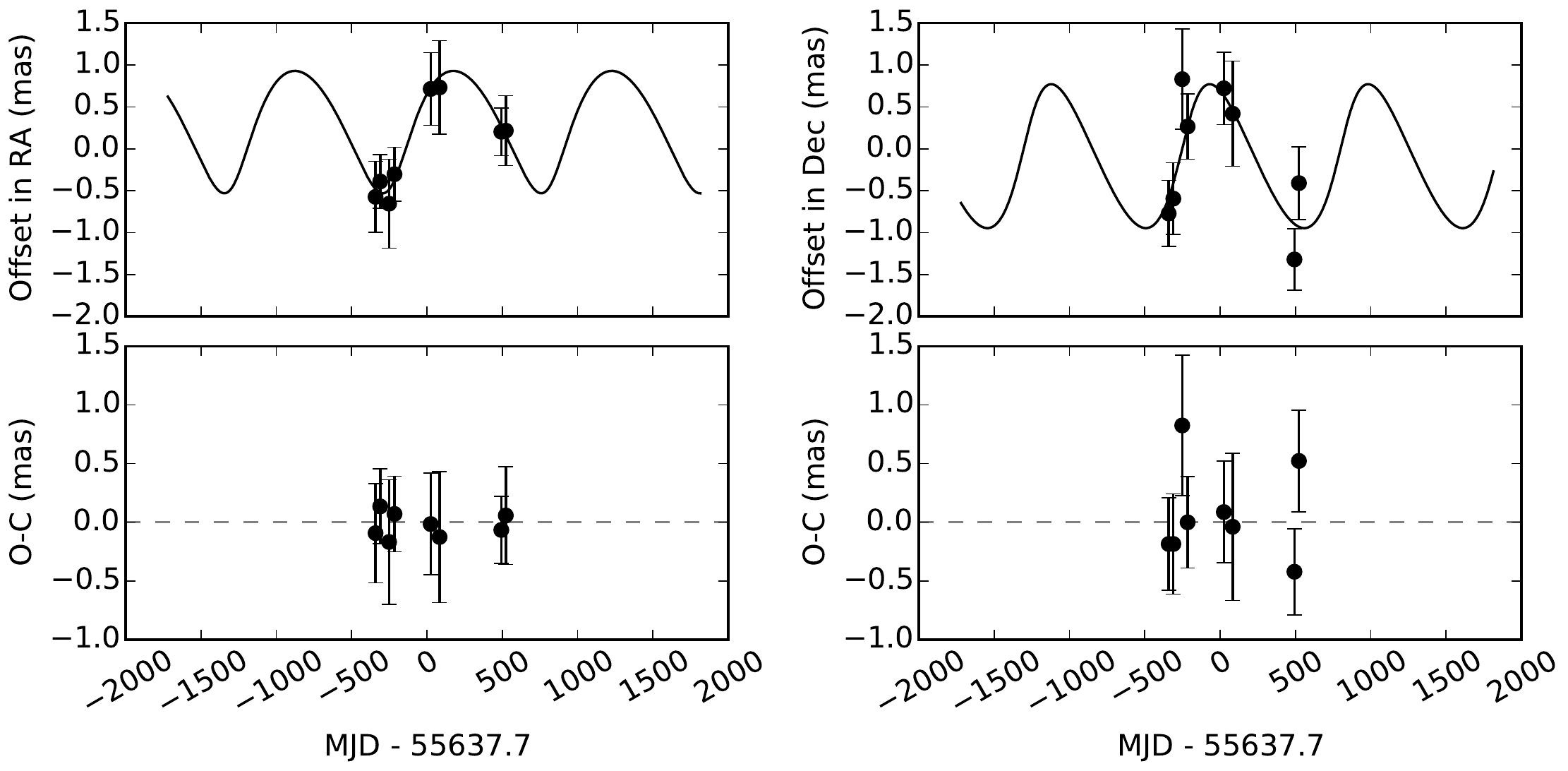}
\caption{Orbital motion of \mystar\ as a function of time. The orbital signature in Right Ascension (\emph{left}) and Declination (\emph{right}) is shown, where black symbols show epoch average values with uncertainties. The bottom panels show the observed minus calculated residuals of epoch averages. FORS2 data cover 82 \% of the orbital period of planet `b`.}
\label{fig:axomc}
\end{figure*}

We conclude that the astrometric orbit of planet `b` was detected with our ground-based astrometry. The adopted parameter values are $\Omega=208^{+15}_{-13}$\degr, $i=45^{+ 21}_{- 11}$\degr, and the semimajor axis of \mystar's reflex motion is $a_1=1.0^{+0.3}_{-0.2}$ mas. This corresponds to a mass of planet \mystar\ b of $M_\mathrm{b}=6.7^{+1.8}_{-1.5}\,M_\mathrm{J}$. These parameters {(and the parallax and proper motions)} are in good agreement with the preliminary values determined from the independent sequential analysis in Sect.\ \ref{sec:astroparam}. The planet mass uncertainty does not account for the uncertainty in the host star mass.

\begin{figure}[h!]
\centering
\includegraphics[width=0.8\linewidth]{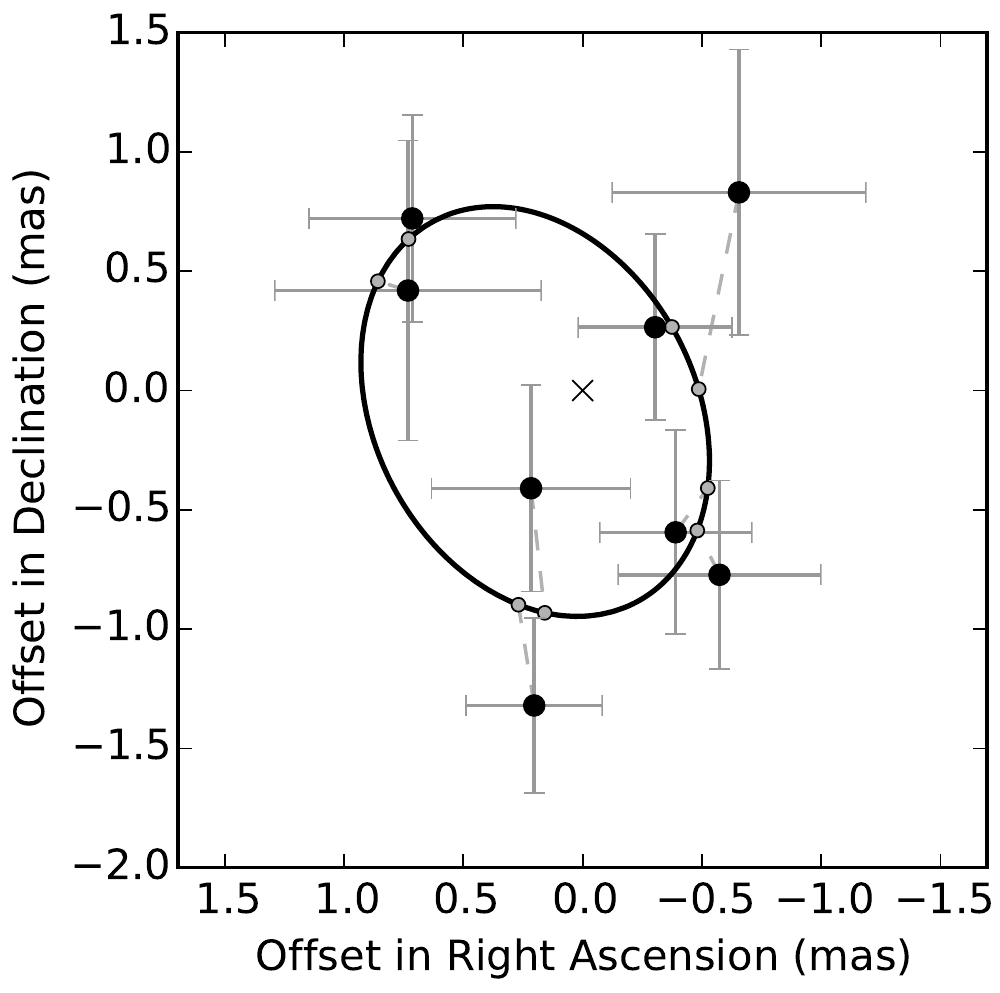}
\caption{Astrometric reflex motion of \mystar\ caused by planet `b` about the system's barycentre (marked by a cross) in the plane of the sky. FORS2 measurements are shown by black circles with uncertainties and dashed grey lines connect to the best-fit model positions (grey circles). North is up and east is left. The motion is counterclockwise.}
\label{fig:ax}
\end{figure}

Figure \ref{fig:axomc} shows the astrometric orbital motion as a function of time. The epoch residuals of the best-fit orbit model have an r.m.s.\ dispersion of 0.28 mas (reduced $\chi^2_\mathrm{epoch}=0.8$), which is almost twice smaller that the residuals of 0.54 mas for the standard model without orbit presented in Sect.\ \ref{sec:astroparam}. This is further strong evidence that the astrometric orbit was detected. The residual dispersion is smaller than the average uncertainty of 0.43 mas, which may indicate that the latter are slightly overestimated. Finally, we present the astrometric orbit of \mystar\ caused by planet `b` in the plane of the sky in Fig.\ \ref{fig:ax}.

\subsection{FORS2 proper motion bias caused by planet `c`}\label{sec:bias}
Because our FORS2 measurements cover only $\sim$12 \% of the orbital period of planet `c`, the corresponding reflex motion of \mystar\ results in a small bias of the measured  proper motion reported in Tables \ref{tab:solution} and \ref{tab:ppm}. To quantify the bias we have to take into account the inclination $i_\mathrm{c}$ and orientation $\Omega_\mathrm{c}$ of planet c's orbit. The distribution of mutual inclinations for long-period giant planets in multi-planetary systems is essentially unknown, because inclinations were constrained in only a few systems \citep{McArthur:2010kx, Fabrycky:2010aa, Correia:2010qq, Pueyo:2015aa}. We conservatively assumed a circular orbit for planet `c` with an inclination that matches the one of planet 'b', which results in a planet mass of $M_c = 9.5\,M_\mathrm{J}$. Furthermore, we assumed that the ascending node $\Omega_\mathrm{c}$ of planet `c` is randomly oriented, which translates into a random distribution of mutual inclinations between `b` and `c`. 

We simulated astrometric data at the FORS2 observation epochs in a Monte Carlo fashion that take into account the orbit of planet `c` and fitted those data with the standard parallax $+$ proper motion model. The difference between the proper motion determined from simulated data with and without planet `c` yields an estimate of the proper motion bias, which amounts to $0.0 \pm1.1$\,mas yr$^{-1}$ in both $\mu_{\alpha^\star}$ and $\mu_{\delta}$. If instead we assume that the planets have both the same inclination and the same ascending node, the corresponding bias is $-0.5 \pm0.3$\,mas yr$^{-1}$ and $-1.4 \pm0.4$\,mas yr$^{-1}$ in $\mu_{\alpha^\star}$ and $\mu_{\delta}$, respectively. {These biases scale with the mass of planet `c` and may slightly increase if its orbit is eccentric.}

\section{Orbital motion of the \mystar\ - \mystarB\  binary}\label{sec:AB}
We measured the proper motion of \mystarB\ relative to \mystar, which is useful to characterise the dynamics of this wide binary system. The proper motion differences  $ \Delta\mu_{\alpha^\star}= \mu_{\alpha^\star} (\textrm{\mystarB})  - \mu_{\alpha^\star} (\textrm{\mystar})$ and $ \Delta\mu_{\delta}= \mu_{\delta} (\textrm {\mystarB})  - \mu_{\delta} (\textrm{\mystar})$  are robust observables, because they are independent of the zero-points of proper motion and of common systematic error components. We performed an astrometric reduction for \mystarB, using the same procedures applied to \mystar\ (Sect.\ \ref{sec:fors2ax}). Table \ref{tab:tbl1} lists the values of $\Delta\mu ^*_{\alpha}$  and $\Delta\mu_{\delta}$ obtained from our data and from three literature catalogues (PPMXL, NOMAD, and UCAC4) that present proper motion values for both stars, which means that those proper motions were determined in the same system, thus mitigating systematic error in the differences $\Delta\mu ^*_{\alpha}$  and $\Delta\mu_{\delta}$.

\begin{table}[h!]
\caption{Relative proper motions of \mystarB\  and  \mystar} 
\centering
\begin{tabular}{@{}ccl@{}}
\hline
\hline
  $\Delta\mu_{\alpha^\star}$ mas/yr  & $ \Delta\mu_{\delta}$ mas/yr        &  Reference\\
  (\hbox{mas yr$^{-1}$}) & (\hbox{mas yr$^{-1}$})      &  \\
  \hline
 -8.73 $\pm 0.63  $   &   1.58  $\pm 0.39   $ &FORS2 (this work) \\
 -17.3 $\pm 8  $         &   4.4  $\pm 8   $        &PPMXL \citep{PPMXL} \\ 
    5.2 $\pm 8  $         &   29.4  $\pm 23   $    & NOMAD \citep{NOMAD} \\
 -33.2 $\pm 9  $         &   -0.5  $\pm 8.5   $    &UCAC4 \citep{UCAC4} \\
\hline
\end{tabular}
\label{tab:tbl1}
\end{table}

The literature relative motions are marginally significant and variable in their signs which leaves the direction of relative motion uncertain. In contrast, our astrometry clearly detects  the relative motion of  \mystarB\  and  \mystar. When investigating the relative orbital motion in this stellar system, we have to account for the influence of the giant planets around \mystar. The proper motion bias due to planet `b` is negligible, because our FORS2 data cover almost a full orbit. However, the bias due to planet `c` that we estimated in Sect.\ \ref{sec:RVanalysis} has the effect of increasing the formal uncertainties of \mystar's proper motion by $ \pm1.1$\,mas yr$^{-1}$ both in RA and Dec. 

To study the orbital configuration of the binary we used the constraints set by the projected distance between the two stars $\Delta \alpha^*$,$\Delta \delta$ and by the relative proper motion $\Delta\mu_{\alpha^\star}$,$\Delta\mu_{\delta}$, which in the plane of the orbit is $8.87 \pm 1.33$\,mas yr$^{-1}$ and which translates into a tangential velocity of $V_{\textrm{t}}=0.704 \pm 0.106$\,\kmps. We  also know \mystar's  systemic velocity $\gamma_0=-39.038$\,\kmps\, (Table \ref{tab:solution}) and we measured the radial velocity of \mystarB\ as $-39.396 \pm 0.003$\,\kmps. The radial velocity of \mystarB\ relative to \mystar\  is then $\Delta \textrm {RV}=-0.358\pm 0.010$\,\kmps, where we accounted for an additional RV uncertainty due to planet c's incomplete orbit and to potential zero-point offsets. Thus, we have constrained all components of the \mystarB's relative velocity vector and of its relative spatial position, except for the distance $z$ to \mystar\ along the line of sight.

The binary's Keplerian motion is described by six parameters: the semimajor axis $a$ of the relative orbit, eccentricity $e$, time of periastron passage, inclination $i$, argument of periastron $\omega$, and ascending node $\Omega$. We assumed a system mass of $1.02 M_{\sun}$, with component masses of $0.71\,M_{\sun}$ and $0.29\,M_{\sun}$ for the stars \citep{Forveille:2011lr} and  $\sim$0.02\,$M_{\sun}$ for the giant planets.

\begin{figure}[h!]
 \centering
\includegraphics[width= \linewidth, trim= 0 0cm 0 0cm,clip=true]{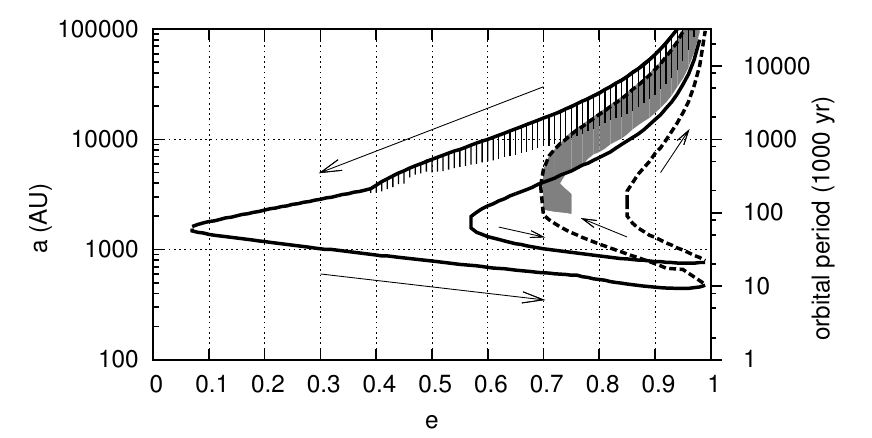}
\caption{Allowed parameter space of semimajor axis $a$ and eccentricity $e$  of the relative orbit of \mystarB\ around \mystar\ at 3-$\sigma$: The area between solid lines corresponds to line-of-sight relative distances $z<z_0$, whereas the area between dashed lines is for $z>z_0$. Arrows show the directions of increasing $z$ and the hatched and filled areas correspond to positions of \mystarB\ in front and behind \mystar, respectively, with relative distance $z$ exceeding $z_{\textrm{Gaia}} = 2400$\,AU that can be resolved with \emph{Gaia} parallaxes.}
\label{ae}
\end{figure}

We ran Monte Carlo simulations to explore the allowed parameter values under the constraints discussed above. The result in the $a-e$ plane is shown in Fig.\,{\ref{ae}}. The allowed parameter space (at 3-$\sigma$ level) is split in two separated domains: for short line-of-sight distances $z<z_0$ between \mystar\ and \mystarB, the allowed parameter space is the area inbetween the solid lines. For large distances $z>z_0$, the allowed area is delimited by dashed lines. The value $z_0 \approx +360$~AU is the particular point where the velocity vector points towards \mystar.  The allowed range of $z$ values is $\pm 7500$~AU, which is unresolved with the  FORS2 astrometry. However, {\it Gaia} may be sensitive to the cases of extreme $z$-values. For $V=7 - 12$ stars, the parallax precision of {\it Gaia}\footnote{\href{http://www.cosmos.esa.int/web/gaia/science-performance}{http://www.cosmos.esa.int/web/gaia/science-performance}} is $\sim$0.01 mas \citep[e.g.][]{de-Bruijne:2012kx} or $\sim$800 AU if expressed in terms of the distance $z$. After solving for the astrometric orbit of \mystar\ due to planet `b`, {\it Gaia} will distinguish at 3-$\sigma$ confidence between solutions with large negative $z<-2400$~AU (hatched area between solid lines in Fig.\,\ref{ae}), large positive $z>2400$~AU (filled area between dashed lines in Fig.\,\ref{ae}), and intermediate $z$ values.

The space of allowed $a-e$ values is sensitive to errors in proper motion, especially in RA, whereas the uncertainties in RV are negligible. With the {\it Gaia} results, we expect that the allowed $a-e$ values will be significantly better constrained, although they will neither resolve the correlation between $a$ and $e$ nor remove the ambiguity of solutions with $z<z_0$ and $z>z_0$. 

Estimates of the angular orbital parameters are strongly affected by the uncertainty in the proper motion bias due to planet 'c' and by the ambiguity in the  $z$ value. However, for moderately eccentric orbits with $e<0.9$ we find  that the inclination is constrained between $50<i<120\degr$.  The allowed ranges for the ascending node are $60<\Omega <130\degr$ if $z>z_0$ and $260<\Omega <300\degr$ if $z<z_0$.

\begin{figure}[h!]
 \centering
\includegraphics[width= \linewidth, trim= 0 0cm 0 0cm,clip=true]{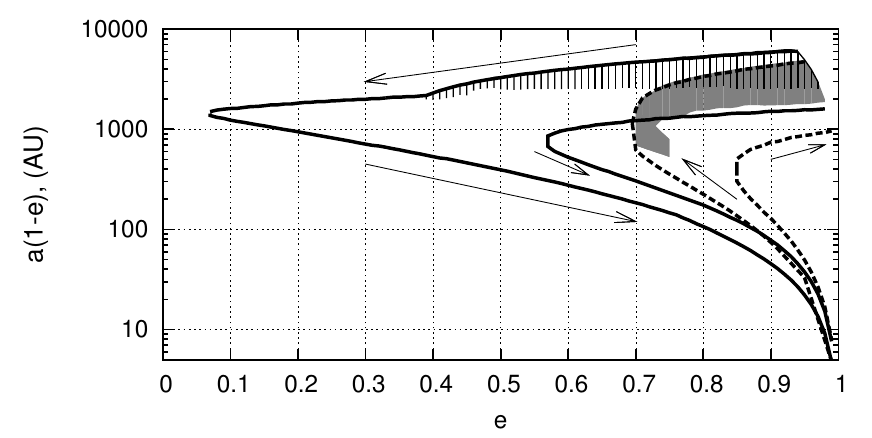}
\caption{Periastron distance $a\,(1-e)$ as a function of eccentricity for the relative binary orbit. The allowed parameter values are indicated like in Fig.\ \ref{ae}.} \label{mind}
\end{figure}

The periastron distance $a\,(1-e)$ of the binary is of particular interest because of its influence on the dynamics of the planetary system around \mystar. We found that its upper limit is 1000--7000 AU almost independently of eccentricity, whereas its minimum value depends strongly on $e$. For nearly circular orbits, it is about 1000 AU and decreases to 100--200~AU for $e \approx 0.8$ and $z<z_0$. For even more eccentric orbits, the periastron distance can be as small as 10~AU, and the limiting value is 5 AU (see Fig.\ \ref{mind}), which also corresponds to the shortest orbital period of $\sim$ 10\,000 years (Fig.\ \ref{ae}). We conclude that the relative distance between \mystarB\ and \mystar\ throughout their orbit most likely remains much larger than the extend of the known planetary system around \mystar, thus the wide binary is unlikely to have affected the formation and evolution of these planets. Additional RV measurements for \mystarB\ and the measurements of the \emph{Gaia} mission will make a better characterisation of the binary orbit possible.

\subsection{Trigonometric parallax of \mystarB}\label{sec:ABpar}
In Sect.\ \ref{sec:fors2ax} we explained that the trigonometric parallax of \mystarB\ was derived under the assumption that it is indistinguishable from that of \mystar, which was necessary to mitigate the effects of the small number of reference stars. Therefore, the parallax $59.3 \pm 0.3$~mas in Table \ref{tab:solution} applies equally to \mystarB, which to our knowledge is the first parallax measurement for this star. Besides, the orbit modelling in this section yields an upper limit of $\pm 7500$~AU for the difference in distances $z$ to these stars. Hence, their parallax difference cannot exceed $\pm 0.128$~mas.

\section{Prospects for directly imaging the gas giants around \mystar}
Obtaining images of extrasolar planets makes it possible to directly measure their luminosities and spectra \citep[e.g.,][]{Lagrange:2010qf, Chilcote:2015aa}. With present-day instrumentation, this technique is mostly limited to young self-luminous giant planets in very long-period orbits ($\gtrsim$10 AU) around nearby young stars \citep[e.g.,][]{Macintosh:2015aa}. The next generation of exoplanet imaging efforts on the ground and in space will target long-period, mature Jovian planets similar to \mystar\,b \citep{Traub:2016JATIS, Kasper:2015conf} and will pave the path towards observing the spectrum of a potential Earth-twin, which will require a large space observatory. Imaging the giant planets around \mystar\ is a stepping stone on this path.

The extreme technical requirements of high-contrast instrumentation, combined with the scientific motivation to prioritise atmospheric characterisation over blind imaging searches, creates a strong incentive to leverage radial-velocity and astrometry programs to cull a target sample with well-constrained ephemeris and masses. Therefore, the task of assessing the observability of indirectly detected exoplanets is already underway, well in advance of the commissioning of such facilities~\citep[e.g.,][]{Howard:2014ExEP, Crossfield:2013aa}. 
 
\subsection{VLT/NaCo observations}
\begin{figure}[h!]
\centering
\includegraphics[width=0.8\linewidth]{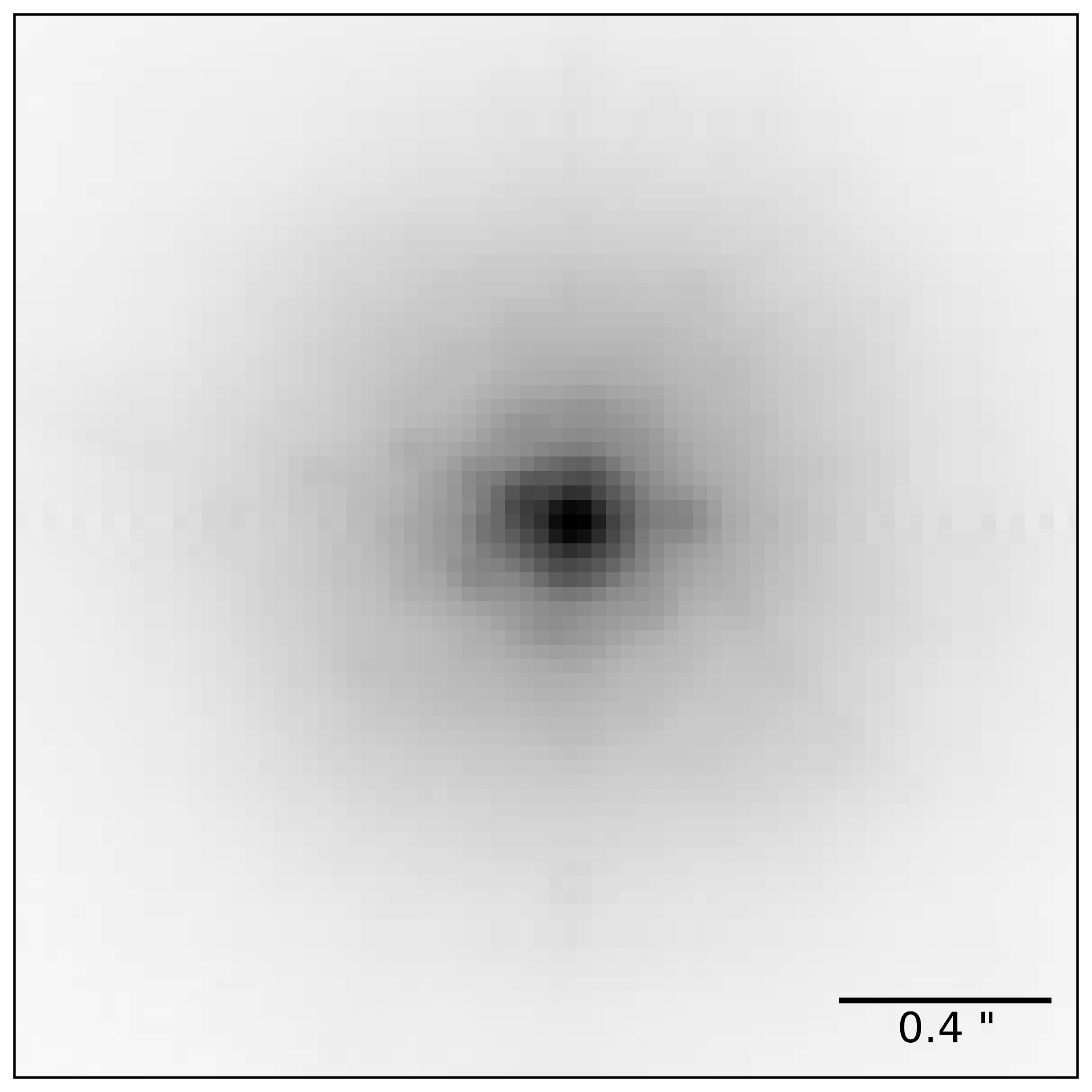}
\caption{NaCo image of \mystar\ in $K_\mathrm{S}$-band.}
\label{fig:nacoimg}
\end{figure}

\begin{figure}[h!]
\centering
\includegraphics[width=\linewidth]{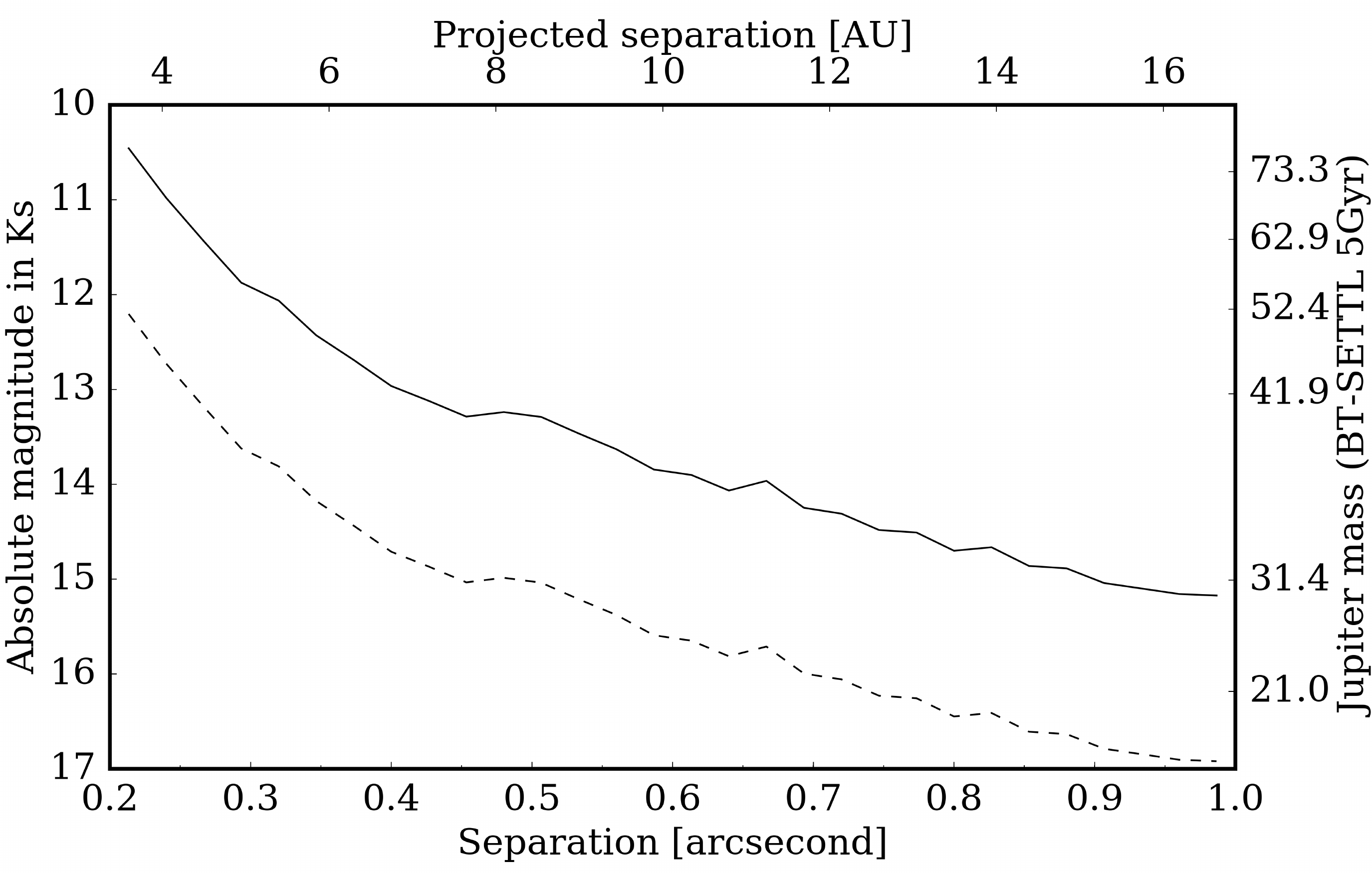}
\caption{Limits on a companion to \mystar\ ($K=5.8$) from the NaCo direct-imaging observations.
The dashed and solid line shows detection limits at 1- and 5-$\sigma$, respectively. The mass estimates (right-hand axis labels) are based on BT-SETTL models at 5 Gyr. Companions above the solid line would have been detected with $>$5-$\sigma$ significance.}
\label{fig:nacoDetLimit}
\end{figure}

\mystar\  was observed on 2010-10-15 with VLT/NaCo \citep{Rousset:2003ys, Lenzen:2003vn} as part of a project aimed at imaging potential sub-stellar companions of red dwarfs that exhibit a radial-velocity drift (ESO Program 086.C-0515(A), PI Montagnier). The 20-minute $K_\mathrm{S}$-band observing sequence was taken in saturated non-coronagraphic field-tracking mode, bracketed by two short non-saturated sequences for photometric calibration. The data were analysed using GRAPHIC \citep{Hagelberg:2016aa} and the reduced image is shown in Fig.\ \ref{fig:nacoimg}. No additional point-source was detected in this snapshot observation. The detection limits shown in Fig. \ref{fig:nacoDetLimit} were derived from the reduced image following a procedure similar to the one described in \cite{Chauvin:2015aa}, and the mass estimates are based on the BT-SETTL CIFIST2011 models \citep{Baraffe:2015aa} using a conservative age estimate of 5 Gyr for the system.

According to the best orbital solution determined in Sect.~\ref{sec:mcmc}, the relative separation of planet `b` at the NaCo observation epoch was 70 mas. This is much smaller than the inner working angle (IWA) of $\sim$200 mas achieved with this dataset, thus NaCo does not yield any constraint on planet `b`. 

For planet `c` in the most conservative case where we assume that its orbit is seen edge-on and $M_c = M_c\sin{i}=6.8\pm0.1\,M_\mathrm{J}$, the relative separation at the NaCo epoch was 170 mas. Figure \ref{fig:nacoDetLimit} shows that at the smallest separation probed with NaCo we can exclude an object with mass $\gtrsim$$75\,M_\mathrm{J}$ with high confidence. On the basis of the NaCo data alone, we can thus exclude that the RV signature corresponding to planet `c` is caused by a stellar companion in an almost face-on orbit, which otherwise would be detectable in the images. Furthermore, the relative separation at the NaCo epoch increases with companion mass. If the orbital inclinations of planets `b` and `c` are equal, in which case $M_c = 9.5\,M_\mathrm{J}$, the relative separation at the NaCo epoch increases to 300 mas and the corresponding upper 5-$\sigma$ mass-limit derived from the images is $\sim$$52\,M_\mathrm{J}$. Following this argument, we can set a lower limit of 10\degr\ to the inclination of planet `c`, because this configuration with $M_c = 39\,M_\mathrm{J}$ and separation 390 mas is ruled out by the NaCo images. The mass of planet `c` has therefore to be in the range of $6.8-39\,M_\mathrm{J}$.

\subsection{Thermal imaging of planets `b` and `c`} 
Even at its minimum mass of 4.9 $M_\mathrm{J}$, \cite{Quanz:2015aa} noted \mystar\,b as a potential target for thermal infrared imaging with the future METIS instrument on the European Extremely Large Telescope (E-ELT). Assuming an age of 5 Gyr, the planet is above the planned sensitivity limits of METIS in the thermal infrared $L$ and $M$ bandpasses. Since the true mass of \mystar\,b established here is yet higher, it is confirmed as an attractive target for ground-based infrared imaging efforts with the E-ELT or other extremely large telescopes with mid-infrared high-contrast instruments.

Moreover, we put forward planet `c` as a new promising target. Its minimum mass is $M_c\sin{i}=6.8\pm0.1\,M_\mathrm{J}$, thus approximately the same as the mass of planet `b`, but \mystar\,c is on a wider orbit with larger relative separation from its host star. Using the estimates for the $6.8\,M_\mathrm{J}$ planet \object{HD\,111232}\,b given in Table 2 of \cite{Quanz:2015aa}, we extrapolated apparent $L$-band magnitudes of $\sim$20.5 and $\lesssim$20.5 and maximum separations of 110 mas and 390 mas for \mystar\,b and \mystar\,c, respectively. Thus both planets can be observed with METIS given the expected limiting magnitude of $L<22.4$ and IWA of 38 mas \citep{Quanz:2015aa}.

We do not expect JWST to image the gas giants around \mystar, due to their small angular separations relative to the inner working angle limits of the various coronagraph modes. In principle, the 0.36\arcsec\ inner working angle of the four quadrant phase mask (FQPM) coronagraph at 11.4 microns enables searching for planets at the same angular separation as planet `c`, with apoastron angular separation 0.4\arcsec\ \citep{Boccaletti:2015aa}. However, the predicted PSF subtraction residuals within 1\arcsec\ of the star will remain too high to reach the required contrast ($\sim$$10^{-5}$) to detect such a cool planet in the mid-infrared. The expected 3--5 $\mu$m planet-to-star contrast ratios of the planets ($\sim$$2\cdot10^{-6}$) are also too extreme for the NIRCam coronagraph detection limits inside 0.5\arcsec\ \citep{Beichman:2010aa} and for NIRISS aperture masking interferometry \citep{artigau2014}.

\subsection{Reflected light observations  with WFIRST} 
We investigated suitability of \mystar\,b and \mystar\,c for observation in reflected starlight with the visible wavelength, space-based coronagraph planned for NASA's Wide-Field Infrared Survey Telescope (WFIRST, \citealt{Noecker:2016JATIS}).

Due to their super-Jovian masses, we expect the planets' internal energy to dominate the stellar irradiation in determining its effective temperature. In the giant-planet evolution models of \citet{Burrows:2004ApJ}, the effective temperatures ($T_{\rm eff}$) of 5 Gyr-old gas giants of masses $6\,M_\mathrm{J}$ and $8\,M_\mathrm{J}$ are 216 K and 251~K, respectively. By comparison, even at planet b's periastron ($\sim$1.24 AU), for an assumed \mystar\ bolometric luminosity of $L_\star = 0.33L_\odot$, and a Bond albedo as low as 0.1, the equilibrium temperature would peak at only 185 K. 

In the $T_{\rm eff}$ range of 200--300 K that evolution models predict for planets `b` and `c`, water condenses in the troposphere, resulting in a higher albedo than Jovian-type (cooler and NH$_3$ cloud-dominated) gas giants \citep{Sudarsky:2000aa}. At a wavelength of 480 nm, the respective geometric albedos of Jupiter and Saturn are 0.46 and 0.39, respectively \citep{Karkoschka:1998Icarus}. In the atmosphere models computed by \citet{Sudarsky:2005ApJ}, a giant planet with $T_{\rm eff}$ similar to \mystar\,b would have a geometric albedo of 0.5 at 500 nm, rising toward 400 nm.

\begin{figure}[h!]
\centering
\includegraphics[width=\linewidth]{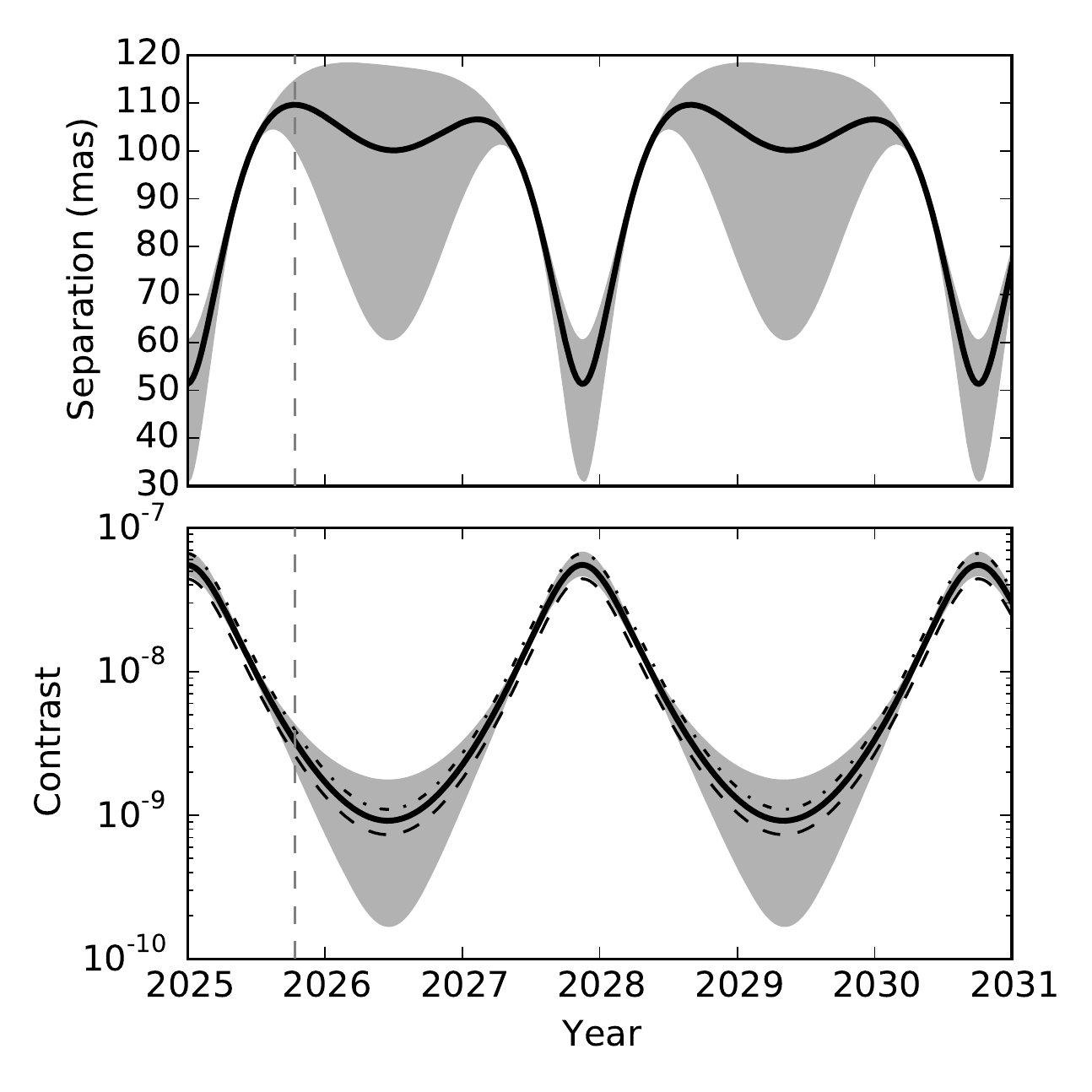}
\caption{Projected separation (top panel) and estimated reflected-light contrast in the WFIRST blue channel (bottom panel) for \mystar\ b as a function of time. The solid line shows the best-fit model and the grey band encompasses the 1-$\sigma$ interval for orbital solutions drawn from the MCMC posterior distribution, i.e.\ it takes into account all uncertainties except for the one in \mystar's mass. In the bottom panel, dashed, solid, and dash-dotted lines represent albedos of 0.4, 0.5, and 0.6, respectively. The vertical dashed line indicates the maximum separation of 110 mas for which we estimated a contrast of $2.6\cdot10^{-9}$ for an albedo of 0.5.}
\label{fig:wfirstConstrast}
\end{figure}

In Fig.\ \ref{fig:wfirstConstrast} we show the estimated contrast of reflected light from planet \mystar\,b in the WFIRST blue channel as a function of time during a 6-year window that may correspond to the WFIRST mission.  We used the Keplerian orbital elements and a Lambert sphere scattering model with a classical phase function to compute the time-evolution of projected angular separation and the planet-to-star contrast ratio, and we assumed that the planets have radii equivalent to Jupiter. The angular separation plot captures the uncertainty in the ephemeris with a 1-$\sigma$ contour of an ensemble of draws from the posterior distributions of the orbital solution. The contrast curve is repeated for three geometric albedos, 0.4, 0.5, and 0.6.

\begin{figure}[h!]
\centering
\includegraphics[width=\linewidth]{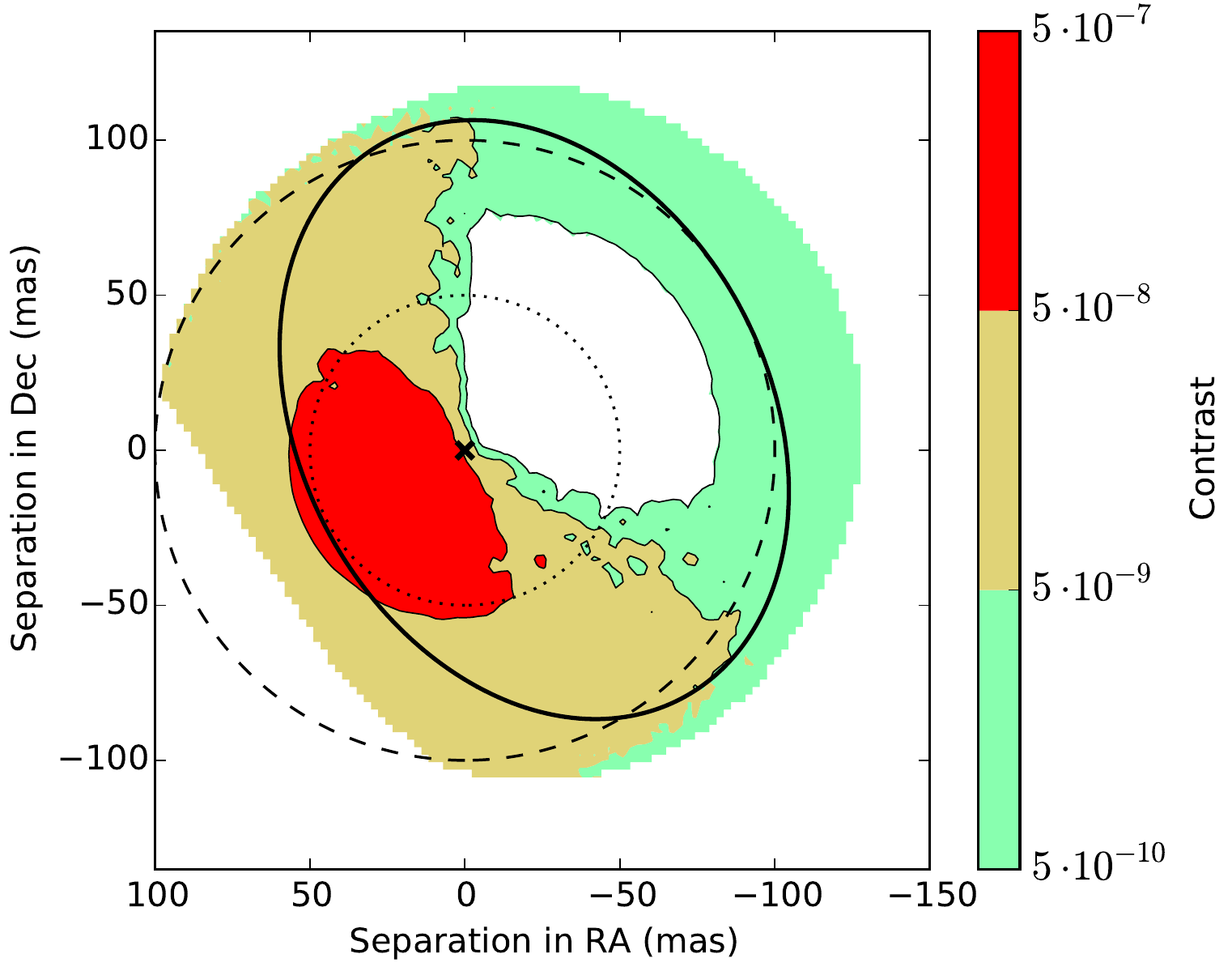}
\caption{Relative position and contrast of \mystar\,b in the WFIRST blue channel projected onto the sky plane for an albedo of 0.5. In combination with Fig.\ \ref{fig:wfirstConstrast}, this allows for optimal planning of direct-imaging observations. The data is the same as the one underlying Fig.\ \ref{fig:wfirstConstrast}. Contrast contours are shown for the ensemble of orbital solutions and the solid line shows the best-fit solution. Dashed and dotted circles indicate separations of 50 mas and 100 mas, respectively, from the host star marked with a cross.}
\label{fig:wfirstContrast2D}
\end{figure}

In Fig.\ \ref{fig:wfirstContrast2D} we show contrast and separation in the sky plane. This representation allows us to define the orientations that corresponds to regions of optimal separation and contrast. In this way, both the epoch and the expected location of planet `b` are defined, which is crucial to optimise the efficiency of these resource-intensive observations. For instance, the instrument can be oriented in a way that the planet falls onto a preferred zone in the image plane.

At a wavelength of $\sim$500\,nm and maximum projected separation of 110 mas, the planet-to-star contrast is $\sim$$2.6\cdot10^{-9}$, above the nominal post-processed detection limits of the WFIRST Coronagraph Instrument \citep{Krist:2015JATIS}. However, this angular separation is just inside the baseline $3.0\,\lambda/D$ inner working angle of the instrument that corresponds to 120 mas at 465 nm, the central wavelength of the bluest imaging bandpass. Therefore, the feasibility of imaging \mystar\ b with WFIRST remains uncertain. Ultimately, it will depend on the final coronagraph mask specifications, which are subject to ongoing design studies trading between throughput, contrast, inner working angle, and bandwidth \citep{Trauger:2016JATIS, Zimmerman:2016JATIS}.

We performed a similar calculation for \mystar\ c using its preliminary ephemeris under the assumption of an edge-on circular orbit. We found that the planet-to-star contrast in reflected light reaches the range of $1\cdot10^{-9}$ -- $2\cdot10^{-9}$ for separations between 200 mas and 390 mas. Because of its larger orbital separation planet `c` receives less insolation and is therefore as challenging as planet `b` at maximum separation in terms of contrast. At the same time its larger projected separation makes it a promising target for WFIRST coronagraphy.

\section{Discussion}
The astrometric measurement of the reflex stellar motion caused by orbiting planets is difficult with current instruments because because the amplitude of the signal is of the of order 1 mas or smaller. The detection of such small signals is significantly eased by targeting stars with giant planets previously characterised with radial velocities.  So far, it has been predominantly achieved from space with the Hubble Space Telescope fine guidance sensor \citep{Benedict:2006kx, Benedict:2010ph, McArthur:2014aa}. In addition, upper limits on the masses of known exoplanets were set with the \emph{Hipparcos} space mission \citep[e.g.][]{Perryman:1996vn, Reffert:2011lr, Sahlmann:2011lr} and in rare cases from the ground \citep{Anglada-Escude:2012vn}.

Here, we demonstrate the detection of the astrometric orbit of an exoplanet host star with a ground-based instrument. Thanks to the outstanding astrometric precision achieved with FORS2/VLT and the detailed knowledge of the orbit's spectroscopic orbital parameters from HARPS RV monitoring, the detection is made with high confidence and we can constrain the inclination and ascending node of \mystar's orbit caused by planet `b`. As a direct consequence of measuring the orbital inclination, we determine the mass of planet `b` to $M_\mathrm{b} = 6.7^{+1.8}_{-1.5}\,M_\mathrm{J}$. 

We measured the minimum mass of the outer giant planet `c` $M_\mathrm{c} \sin i = 6.8\pm0.1\,M_\mathrm{J}$ under the preliminary assumption of a circular orbit and constrained its inclination to be larger then 10\degr. We do not set observational constraints on the orbital inclinations of the inner planets, but if the system is aligned, i.e.\ all planets share similar inclinations like the planets in the Solar System, the masses of planets `c`, `d`, and `e` are $\sim$40\,\% higher than their minimum masses derived from RV. In the future, the determination of the mutual inclinations between the planets using astrometry or other techniques may hint on the dynamical history to the system.

\section{Conclusions}
We pursued the detailed characterisation of the planetary system around the M0 dwarf \mystar\ using a wide range of observational techniques (astrometry, radial velocity, high-contrast imaging) and instruments (FORS2, HARPS, NaCo, \emph{Hipparcos}), which lead to the determination of the mass of planet `b` and the preliminary measurement of the minimum mass of planet `c`. We confirmed the presence of the inner planet `d` and of the periodic signal associated with planet `e`, whose period is close to the star's rotation period. We also find tentative evidence for an additional periodic RV signal at $\sim$1600 days.

We demonstrated how astrometry can leverage radial-velocity planet searches to identify the best targets for future high-contrast direct-imaging observations. The determination of the astrometric orbit yields the comprehensive ephemeris of the planet-star system, which is crucial for efficient planning of the observation timing and the setup/orientation of the instrument. \emph{Gaia}'s astrometry will make it possible to extend these efforts to many other stars with already-known or newly-discovered giant planets. We showed that the outer giant planets around \mystar\ are promising targets for direct imaging of their thermal radiation and/or reflected light with future facilities, in particular extremely large ground-based telescopes like the E-ELT and space missions like WFIRST.

At a distance of 16.7 pc, \mystar\ and its rich planetary system with small inner planets, giant outer planets, and a wide binary companion, represents a fascinating outcome of the star and planet formation process.

\begin{acknowledgements}
J.S.\ is supported by an ESA Research Fellowship in Space Science. J.H.\ is supported by the Swiss National Science Foundation (SNSF). X.B., X.D. and T. F. acknowledge the support of the French Agence Nationale de la Recherche (ANR), under the program ANR-12-BS05-0012 Exo-atmos, as well as funding from the European Research Council under the ERC Grant Agreement n. 337591-ExTrA. This research made use of the databases at the Centre de Donn\'ees astronomiques de Strasbourg (\url{http://cds.u-strasbg.fr}), NASA's Astrophysics Data System Service (\url{http://adsabs.harvard.edu/abstract\_service.html}), the paper repositories at arXiv, of APLPY \citep{Robitaille:2012aa}, and of ASTROPY, a community-developed core Python package for Astronomy \citep{Astropy-Collaboration:2013aa}. Some figures were produced with \texttt{corner.py} (\href{http://dx.doi.org/10.5281/zenodo.45906}{http://dx.doi.org/10.5281/zenodo.45906}). The authors also made use of SCIPY \citep{Jones:2001aa}, NUMPY \citep{Oliphant2007}, IPYTHON \citep{Perez2007}, and MATPLOTLIB \citep{hunter2007}.
\end{acknowledgements}

\bibliographystyle{aa} 
\bibliography{/Users/jsahlmann/astro/papers} 
\begin{appendix} 

\section{Figures and tables}
\begin{figure*}
\centering
\includegraphics[width=\linewidth]{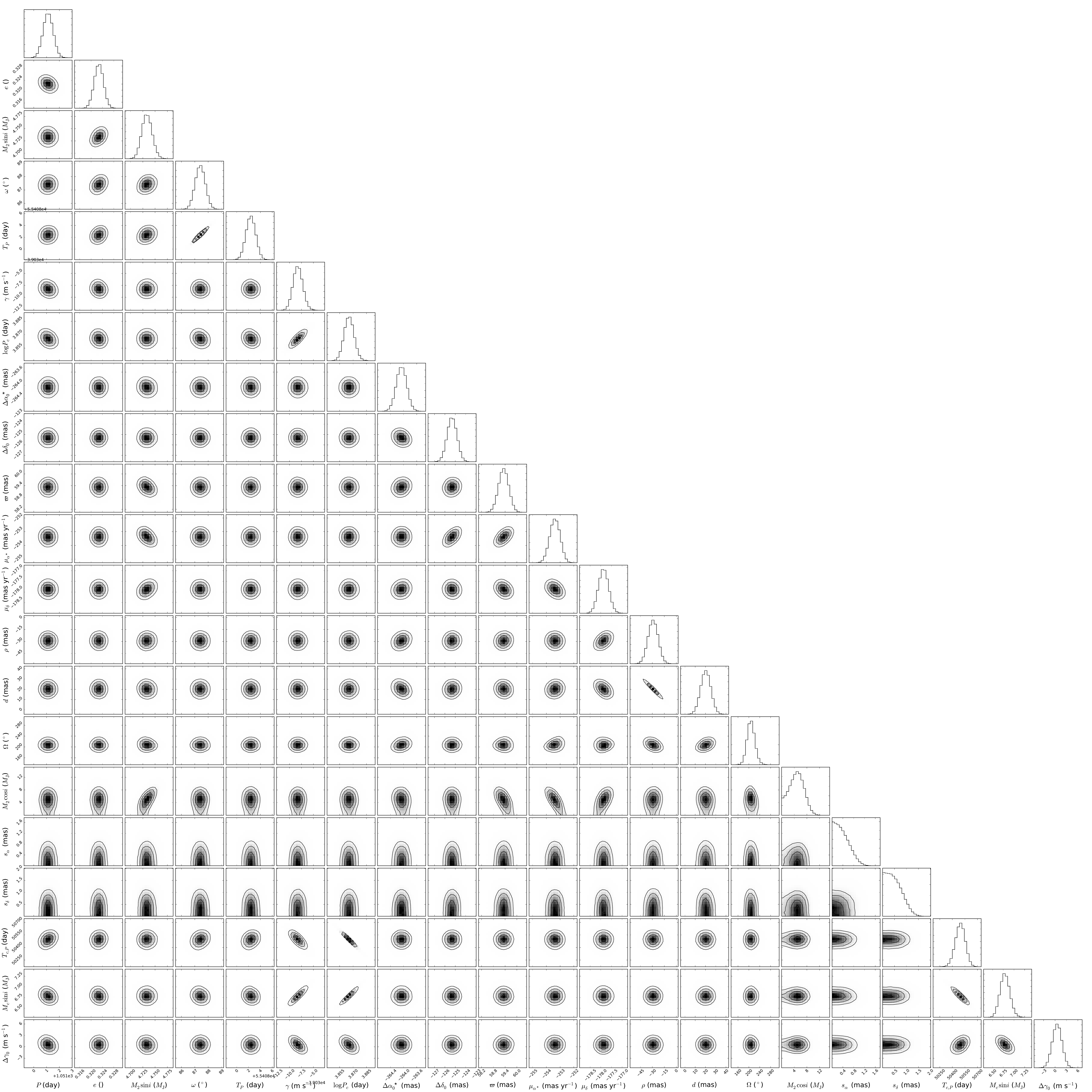}
\caption{Joint marginal distributions of the 21 free parameters probed by the MCMC. The circular shapes of most joint distributions and the quasi-Gaussian marginal distributions of the fit parameters indicate that the model is well constrained. The only strong correlations are present as expected between periastron time ($T_\mathrm{P}$) and argument ($\omega$) for both planets and between the DCR parameters $\rho$ and $d$.}
\label{fig:corner}
\end{figure*}

\begin{table}[h!]
\caption{{Epoch astrometry of \mystar\ in the ICRF, after the effects of DCR have been removed. The quoted uncertainties correspond to the photocentre precision. The conversion to ICRF was done using reference stars catalogued in USNO-B1 \citep{Monet:2003rt} as described in \cite{Sahlmann:2016aa} and introduces an additional uncertainty of 90 mas in both RA and Dec, which was not incorporated here. Note that the analysis in Sect.\ \ref{sec:mcmc} was performed on the basis of individual-frame astrometry.}}
\centering
\begin{tabular}{ccccc}
\hline\hline
Epoch & RA & $\sigma_\mathrm{RA}$ & Dec & $\sigma_\mathrm{Dec}$ \\
(MJD) & (deg) & (mas)& (deg) & (mas)\\
\hline
55296.3070 & 262.54563132 & 0.31 & -51.63746885 & 0.46 \\
55327.2710 & 262.54561268 & 0.30 & -51.63747582 & 0.37 \\
55386.1120 & 262.54556902 & 0.59 & -51.63748333 & 0.68 \\
55422.0060 & 262.54554604 & 0.25 & -51.63748493 & 0.42 \\
55662.3820 & 262.54551810 & 0.32 & -51.63751807 & 0.47 \\
55721.1670 & 262.54547883 & 0.64 & -51.63752977 & 0.78 \\
56131.1510 & 262.54533245 & 0.26 & -51.63758361 & 0.32 \\
56161.0270 & 262.54531500 & 0.33 & -51.63758406 & 0.42 \\
\hline
\end{tabular}
\label{tab:epastr}
\end{table}

\clearpage
\onecolumn

\longtab{
\begin{longtable}{rrr}
\caption{Radial-velocity measurements and uncertainties for \mystar.}\\
\hline \hline
JD-2\,400\,000 & RV & Uncertainty \\
 & (\kmps) & (\kmps) \\
\hline
\endfirsthead
\caption{Continued.}\\
\hline \hline
JD-2\,400\,000 & RV & Uncertainty \\
 & (\kmps) & (\kmps) \\
\hline

\endhead
\hline
\endfoot
\hline
\endlastfoot
53917.747997 & -39.102320 & 0.001460 \\
53919.735174 & -39.095520 & 0.001770 \\
54167.897856 & -39.003570 & 0.001240 \\
54169.895854 & -39.005830 & 0.001160 \\
54171.904445 & -39.002170 & 0.001230 \\
54232.818013 & -39.003590 & 0.001120 \\
54391.491808 & -39.165870 & 0.001110 \\
54393.489934 & -39.170020 & 0.001190 \\
54529.900847 & -39.246180 & 0.001300 \\
54547.915016 & -39.243380 & 0.001130 \\
54559.815698 & -39.236460 & 0.001260 \\
54569.903637 & -39.243920 & 0.001400 \\
54571.889460 & -39.244430 & 0.000950 \\
54582.820292 & -39.235400 & 0.001160 \\
54618.755585 & -39.225970 & 0.001490 \\
54660.661636 & -39.202780 & 0.001110 \\
54661.772229 & -39.204040 & 0.001070 \\
54662.675237 & -39.207460 & 0.001280 \\
54663.811590 & -39.204150 & 0.001020 \\
54664.790043 & -39.199740 & 0.001380 \\
54665.786377 & -39.204870 & 0.001030 \\
54666.696058 & -39.207140 & 0.000970 \\
54670.672602 & -39.204100 & 0.001360 \\
54671.603329 & -39.203460 & 0.001230 \\
54687.561959 & -39.202170 & 0.001240 \\
54721.554874 & -39.185860 & 0.001350 \\
54751.490690 & -39.173490 & 0.002450 \\
54773.502375 & -39.161440 & 0.001070 \\
54916.819805 & -39.098820 & 0.000880 \\
54921.892971 & -39.101670 & 0.001380 \\
54930.906849 & -39.094560 & 0.001190 \\
54931.795103 & -39.093950 & 0.001210 \\
54935.817789 & -39.092340 & 0.000840 \\
55013.686615 & -39.051590 & 0.001200 \\
55013.743720 & -39.052850 & 0.001380 \\
55074.520060 & -39.025660 & 0.001200 \\
55090.507026 & -39.023070 & 0.001180 \\
55091.528800 & -39.023920 & 0.002800 \\
55098.494144 & -39.024050 & 0.000860 \\
55100.540947 & -39.018950 & 0.000980 \\
55101.490472 & -39.018340 & 0.001380 \\
55102.502862 & -39.018330 & 0.001840 \\
55104.540258 & -39.017120 & 0.001310 \\
55105.523635 & -39.018660 & 0.002310 \\
55106.519974 & -39.016990 & 0.001240 \\
55111.509339 & -39.014890 & 0.000980 \\
55113.497880 & -39.015140 & 0.001000 \\
55115.514997 & -39.008440 & 0.002250 \\
55116.487535 & -39.012070 & 0.000950 \\
55117.493046 & -39.008940 & 0.001340 \\
55121.526645 & -39.003150 & 0.001360 \\
55122.505321 & -39.006700 & 0.001260 \\
55124.497834 & -39.006190 & 0.000880 \\
55127.516794 & -39.005430 & 0.000870 \\
55128.513957 & -39.001360 & 0.000870 \\
55129.495404 & -39.002290 & 0.000930 \\
55132.495755 & -39.001630 & 0.000990 \\
55133.493189 & -39.002840 & 0.001060 \\
55259.907275 & -38.962450 & 0.001290 \\
55260.864406 & -38.962750 & 0.001130 \\
55284.893135 & -38.968410 & 0.001710 \\
55340.708504 & -38.986180 & 0.001300 \\
55355.795443 & -39.002450 & 0.001230 \\
55375.610729 & -39.026920 & 0.001410 \\
55387.656686 & -39.041770 & 0.001570 \\
55396.537980 & -39.053040 & 0.001440 \\
55400.642866 & -39.061490 & 0.000980 \\
55401.594785 & -39.057630 & 0.001200 \\
55402.590925 & -39.051980 & 0.003530 \\
55402.702771 & -39.056950 & 0.002280 \\
55404.625157 & -39.064340 & 0.001330 \\
55404.645561 & -39.064370 & 0.002180 \\
55407.576762 & -39.068750 & 0.001390 \\
55424.575448 & -39.094020 & 0.002440 \\
55437.618432 & -39.109210 & 0.001200 \\
55439.623950 & -39.112040 & 0.000970 \\
55443.607470 & -39.124790 & 0.001150 \\
55648.906952 & -39.181170 & 0.001020 \\
55662.922210 & -39.173760 & 0.001040 \\
55711.719077 & -39.150020 & 0.002280 \\
55783.584250 & -39.118190 & 0.001020 \\
55816.541186 & -39.105710 & 0.001050 \\
55839.519405 & -39.097140 & 0.001070 \\
55988.884316 & -39.019810 & 0.000840 \\
56023.833004 & -39.011010 & 0.001060 \\
56115.525064 & -38.957550 & 0.001020 \\
56116.514733 & -38.961190 & 0.001090 \\
56117.503694 & -38.957390 & 0.001060 \\
56118.512585 & -38.957510 & 0.001020 \\
56119.521843 & -38.957700 & 0.001170 \\
56120.664105 & -38.959460 & 0.001300 \\
56121.779723 & -38.952800 & 0.001540 \\
56149.649734 & -38.944460 & 0.001840 \\
56150.522254 & -38.944480 & 0.001080 \\
56152.682008 & -38.943540 & 0.001940 \\
56154.469992 & -38.939200 & 0.001010 \\
56160.486828 & -38.940390 & 0.000970 \\
56161.667048 & -38.937340 & 0.001160 \\
56166.514404 & -38.939610 & 0.001030 \\
56167.501059 & -38.936100 & 0.001130 \\
56168.522469 & -38.936110 & 0.001210 \\
56171.486378 & -38.936310 & 0.000990 \\
56172.506593 & -38.934650 & 0.001020 \\
56175.612974 & -38.930530 & 0.001160 \\
56181.493719 & -38.929480 & 0.001100 \\
56182.536592 & -38.931390 & 0.001200 \\
56186.568383 & -38.925520 & 0.001090 \\
56187.535729 & -38.920520 & 0.001450 \\
56188.542783 & -38.924280 & 0.003940 \\
56190.544160 & -38.924690 & 0.001200 \\
56191.502942 & -38.923340 & 0.000960 \\
56192.535827 & -38.926330 & 0.001040 \\
56193.540694 & -38.925740 & 0.000960 \\
56194.538339 & -38.923360 & 0.001000 \\
56195.550069 & -38.926200 & 0.001110 \\
56196.545087 & -38.922840 & 0.001130 \\
56198.542416 & -38.916380 & 0.001450 \\
56199.548846 & -38.918960 & 0.001940 \\
56201.540471 & -38.915850 & 0.001370 \\
56202.558892 & -38.915170 & 0.002090 \\
56358.902903 & -38.887580 & 0.006420 \\
56364.874908 & -38.893990 & 0.001650 \\
56365.901491 & -38.899390 & 0.000970 \\
56496.645686 & -39.042710 & 0.001240 \\
56805.867212 & -39.066360 & 0.001150 \\
56931.529582 & -38.999420 & 0.001360 \\
57148.820184 & -38.903230 & 0.001510 \\
57434.865699\tablefootmark{a} & -38.847620 & 0.002140 \\
57452.818633\tablefootmark{a} & -38.864550 & 0.001520 \\
\label{tab:rv}
\end{longtable}
\tablefoot{{a} Data taken after the HARPS upgrade.}
}

\end{appendix}

\end{document}